\documentclass[a4paper,11pt]{article}
\pdfoutput=1 
             
\usepackage{jheppub} 
                     
\pdfoutput=1
\usepackage{graphicx,array}
\usepackage{color}
\usepackage{setspace}
\usepackage{amstext}
\usepackage{amssymb}
\usepackage{slashed}
\usepackage{makecell}
\usepackage{multirow}
\usepackage{wasysym}
\usepackage{physics}
\usepackage[T1]{fontenc} 
\usepackage{changepage}
\usepackage{arydshln}
\usepackage{tikz}
\usetikzlibrary{tikzmark}           
\usepackage{scalerel} 
\usepackage{slashed}
\usepackage{dsfont}

\definecolor{orange}{rgb}{1,0.5,0}
\definecolor{brown}{rgb}{0.59, 0.29, 0.0}

\definecolor{note_fontcolor}{rgb}{0.80078125, 0.80078125, 0.80078125}

\definecolor{darkgreen}{rgb}{0,0.5,0}

\def\beq{\begin{equation}}
\def\eeq{\end{equation}}
\def\bea{\begin{eqnarray}}
\def\eea{\end{eqnarray}}

\newcommand{\eq}[1]{\begin{align}\begin{split}#1\end{split}\end{align}}

\def\IZ{{\mathbb{Z}}}
\def\IQ{{\mathbb{Q}}}
\def\IR{{\mathbb{R}}}

\def\CA{{\cal A}}
\def\CB{{\cal B}}
\def\CC {{\cal C}}
\def\CD {{\cal D}}

\def\CG {{\cal G}}
\def\CH {{\cal H}}

\def\CO {{\cal O}}

\def\CO {{\cal O}}

\def\CG {{\cal G}}
\def\CH {{\cal H}}

\def\CB {{\cal B}}

\def\CS {{\cal S}}

\def\half{\frac{1}{2}}

\newcommand{\centeron}[2]{{\setbox0=\hbox{#1}\setbox1=\hbox{#2}\ifdim
		\wd1>\wd0\kern.5\wd1\kern-.5\wd0\fi \copy0
		\kern-.5\wd0\kern-.5\wd1\copy1\ifdim\wd0>\wd1
		\kern.5\wd0\kern-.5\wd1\fi}}
\newcommand{\ltap}{\>\centeron{\raise.35ex\hbox{$<$}}
	{\lower.65ex\hbox{$\sim$}}\>}
\newcommand{\gtap}{\>\centeron{\raise.35ex\hbox{$>$}}
	{\lower.65ex\hbox{$\sim$}}\>}

\makeatletter
\newcommand*{\Relbarfill@}{\arrowfill@\Relbar\Relbar\Relbar}
\newcommand*{\xeq}[2][]{\ext@arrow 0055\Relbarfill@{#1}{#2}}
\makeatother

\usepackage{cleveref}

\title{\boldmath 
Coupling a Cosmic String to a TQFT}

\author[a,b,c]{T.~Daniel Brennan,}
\author[c,d,e,f]{Sungwoo Hong,}
\author[c,d,g]{Lian-Tao Wang}

\affiliation[a]{Department of Physics, University of California, San Diego}
\affiliation[b]{Kadanoff Center for Theoretical Physics $\&$ Enrico Fermi Institute, University of Chicago, Michelson Center for Physics, 933 E 56th St, Chicago, IL 60637, USA}
\affiliation[c]{Department of Physics, The University of Chicago, Chicago, IL 60637 , USA }
\affiliation[d]{Enrico Fermi Institute, University of Chicago, Chicago, Illinois 60637, USA}
\affiliation[e]{Argonne National Laboratory, Lemont, IL 60439, USA}
\affiliation[f]{Department of Physics, KAIST, Daejeon, 34141, Korea}
\affiliation[g]{Kavli Institute for Cosmological Physics, University of Chicago, Chicago, Illinois 60637, USA}

\abstract{A common framework of particle physics consists of two sectors of particles, such as the Standard Model and a dark sector, with some interaction between them. In this work, we initiate the study of a qualitatively different setup in which one of the sectors is a topological quantum field theory (TQFT). Instead of particles, the physics of a TQFT  only manifests itself in non-trivial spacetime topologies or in the presence of topological defects. In particular, we consider two possible ways in which axionic cosmic strings can interact with a $\mathbb{Z}_n$ TQFT. One of them,  by extending the structure of the axion coupling, leads to specific predictions for the localized degrees of freedom on the cosmic string, which can in turn effect their evolution and leave observable signals. The second approach, by gauging a discrete subgroup of the axionic shift symmetry, leads to dramatic changes in the string spectrum. We stress that the scenario considered here should be regarded as a plausible way for new physics to arise since it can be the low energy effective field theory for quite generic scenarios at high energies. To demonstrate this point and further illustrate the physical implications, we construct UV completions for both of the cases of couplings to TQFTs. While detailed predictions for observable signals of such scenarios require further investigation, our results demonstrate that there are rich new phenomena in this scenario.  }

\begin{document}
\maketitle
\flushbottom

\section{Introduction}
\label{sec:Intro}

There are many proposed scenarios of physics beyond the Standard Model. One universally adopted framework to incorporate new physics is to couple a known particle physics sector, such as the full Standard Model, to a ``new physics'' sector which includes \emph{new particles}. These extra local degrees of freedom together with \emph{new interactions}  may introduce novel dynamics and lead to solutions to existing problems in particle physics. %
In studying these new theories, \emph{symmetry} provides an extremely powerful tool. 

Historically, new understandings of symmetry in physics have almost always led to clarifications of existing puzzles and provided new insights. 
In this paper, we initiate the study of a new class of couplings and analyze them by means of new symmetries. Specifically, we study the effect of coupling a particle physics theory described by a local relativistic quantum field theory (QFT) to a topological quantum field theory (TQFT). This idea of ``coupling a QFT to a TQFT'' was introduced in \cite{Kapustin:2014gua} (see also \cite{Seiberg:2010qd, Gaiotto:2014kfa,Pantev:2005rh,Pantev:2005wj,Pantev:2005zs} ).  Yet, to the best of our knowledge,  our current work is the first in considering such possibilities in the context of particle physics.\footnote{For other investigations of TQFT's in phenomenological settings, see \cite{Alford:1989ch,Preskill:1990bm,Alford:1992yx,Wang:2020xyo, Wang:2020mra}.  } Since TQFT is not characterized by any local excitations, understanding the physics of TQFT and TQFT-couplings requires a new set of tools which is afforded by generalized global symmetry \cite{Gaiotto:2014kfa}. 

Our goal is to demonstrate through a couple of simple examples that such TQFT-couplings can lead to non-trivial and interesting phenomenological implications, including possible observable effects, which are  difficult to analyze using traditional methods. We also hope to emphasize that TQFT-couplings, which may appear somewhat exotic, in fact can arise as \emph{discrete IR remnants} in familiar local QFTs. For instance, as described in detail in Appendix~\ref{app:review_BF}, a $U(1)$ gauge theory with a charge $n$ scalar field can flow to a $\mathbb{Z}_n$ TQFT with non-trivial physical observables. This suggests that many theories in particle physics may also have secret TQFT-couplings that arise as discrete remnants which differentiate between  UV completions. In these cases, it is crucial to be able to identify their physical implications and formulate appropriate experimental search strategies. We hope to demonstrate  that using the techniques of generalized global symmetries may provide a path to accomplish this goal. 

Ever since the notion of generalized global symmetry was introduced \cite{Gaiotto:2014kfa}, it has been a very active field of research and has lead to many insights in theoretical QFT (see \cite{Cordova:2022ruw} for a summary and references there-in). Accordingly, it has become increasingly important to determine how to effectively implement generalized global symmetry in the study of particle physics (for example see recent works \cite{Brennan:2020ehu,Anber:2021upc, Cordova:2022fhg, McNamara:2022lrw, Cordova:2022qtz}).


In this paper we will apply the techniques of generalized global symmetries to study the effects of 
coupling a TQFT to axion-Maxwell theory\footnote{In recent particle physics literature, the term axion often refers to the particle associated with the Peccei-Quinn (PQ) solution to the strong CP problem. In this paper, for convenience, we will use axion to denote a general axion-like particle without necessarily assuming it plays a role in the PQ mechanism. With a similar abuse of language, we will also call the associated global symmetry $U(1)_{\rm PQ}$.   }. Axion-Maxwell theory is described by an action\footnote{In the rest of our paper, we will adopt differential form notation. The action of axion-Maxwell theory in terms of differential forms is given by eq.~(\ref{eq:S_0}).}
\beq
S_{\scriptscriptstyle a\text{-}{\rm MW}} = \int \frac{1}{2} \partial_\mu a \partial^\mu a - \frac{1}{4 g^2} F_{A\mu\nu} F_A^{\mu\nu} - \frac{i K_A}{16\pi^2} \frac{a}{f_a} F_A^{\mu\nu} \tilde{F}_{A\mu\nu}
\label{eq:intro_aMW}
\eeq
and appears frequently in the literature. Here, $K_A \in \mathbb{Z}$ is a discrete coupling constant that matches the $U(1)_{\rm PQ} \left[ U(1) \right]^2$ Adler-Bell-Jackiw (ABJ) \cite{Adler:1969gk, Bell:1969ts} anomaly coefficient of any UV completion -- we will have in mind a completion by a KSVZ-type theory~\cite{Kim:1979if, Shifman:1979if}. Here we will distinguish between the axion-Maxwell sector and TQFT sector  by using a subscript $A$ for axion-Maxwell sector  and $B$ for the TQFT sector.

For the TQFT sector, we consider a gauge theory associated with a $\mathbb{Z}_n$ discrete gauge group whose action is given by\footnote{BF theory admits several different descriptions and details can be found in \cite{Kapustin:2014gua}.}
\beq
S_{\scriptscriptstyle \rm BF} =\frac{i n }{2\pi} \int  B^{(2)} \wedge F_B^{(2)} = \frac{in}{4\pi}\int d^4 x \, \epsilon^{\mu\nu\rho\sigma} B_{\mu\nu}^{(2)} \partial_\rho B_\sigma^{(1)}
\label{eq:intro_BF}
\eeq
where, $B^{(2)}$ is a 2-form gauge field (hence two antisymmetric indices) and $F^{(2)} _B= d B^{(1)}$ is the field strength of a 1-form gauge field associated with a gauge group $U(1)_B$  which is restricted to $\mathbb{Z}_n\subset U(1)_B$ by the form of the above action. A review of generalized global symmetry is presented in Appendix~\ref{app:review_GGS} and a detailed discussion of this $\mathbb{Z}_n$ gauge theory can be found in Appendix~\ref{app:review_BF}. As we mentioned already, this TQFT can arise as the IR limit of the Higgs phase of an abelian Higgs model with a charge $n$ Higgs field.

There are many ways to couple axion-Maxwell sector to a $\mathbb{Z}_n$ TQFT, each of which lead to distinct physical effects. In Section~\ref{sec:TQFT-coupling I}, we discuss the TQFT-coupling via axion-portal given by
\begin{eqnarray}
S_{\scriptscriptstyle \rm TQFT\text{-}coupling \; I} = - \frac{i K_{AB}}{4 \pi^2 f_a} \int a F_A^{(2)} \wedge F_B^{(2)} - \frac{i K_{B}}{8\pi^2 f_a} \int a F_B^{(2)} \wedge F_B^{(2)} ~.
\label{eq:intro_couplingI}
\end{eqnarray}
For this coupling, we show that
\begin{itemize}
\item $\left[ \right.$Section~\ref{subsec:anomaly inflow}$\left. \right]$ On an axion string, there must exist a set of chiral modes to cancel gauge anomalies from the bulk topological interactions (via anomaly inflow). In particular, the couplings to the TQFT sector implies that these modes must carry $\mathbb{Z}_n$ charges as well as $U(1)_A$ charges.
\item $\left[ \right.$Section~\ref{subsec:UV model}$\left. \right]$ This theory can be  an IR effective field theory of a extended version of KSVZ theory where the KSVZ fermions are charged under an additional $U(1)_B$ that is spontaneously broken $U(1)_B\mapsto \IZ_n$. 
\item $\left[ \right.$Appendix~\ref{app:symmetry_TQFT-coupling I}$\left. \right]$ The TQFT-coupling in \eqref{eq:intro_couplingI} enriches the generalized global symmetry structure and in particular, modifies the 3-group global symmetry structure. The non-trivial 3-group symmetry implies constraints on the renormalization group flow in the form of (parametric) inequalities among symmetry emerging energy scales of different higher-form symmetries.\footnote{For earlier works on 3-group global symmetries of axion-Maxwell theory, see \cite{Hidaka:2020iaz, Hidaka:2020izy, Brennan:2020ehu}. In addition, recently, it was pointed out in \cite{Cordova:2022ieu} that the axion-Maxwell theory also has an infinite set of non-invertible symmetries \cite{Kaidi:2021xfk, Choi:2021kmx}.}
\item $\left[ \right.$Section~\ref{sec:TQFT-coupling I} and \ref{sec:pheno}$\left. \right]$ These couplings may lead to multiple interesting observable effects. Our theory predicts existence of strings coming from the TQFT sector in addition to the standard axion strings. This may lead to ``coaxial hybrid strings'' whose properties (and existence) have not been studied. 
The fact that the chiral  modes living in the string core have $\mathbb{Z}_n$ charges may additionally lead to significant changes in the evolution of the axion string. 
This in turn might have far reaching consequences such as for ``cosmological plasma collider'' effects and vorton stability.
\end{itemize}

In Section~\ref{sec:TQFT-coupling II}, we discuss a different TQFT-coupling which can be obtained by gauging a discrete symmetry. Concretely, we describe gauging a $\IZ_n$ subgroup of the $\IZ_{K_A}$ (0-form) axion shift symmetry, $a \to a + \frac{2\pi f_a}{K_A}$. This leads to a TQFT-coupling of the form
\beq
S_{\scriptscriptstyle \rm TQFT\text{-}coupling \; II} = \frac{1}{2} \int  (d a - f_a C^{(1)}) \wedge * ( d a - f_a C^{(1)} ) + \frac{iK_A}{f_a} \int (d a - f_a C^{(1)}) \wedge \omega_3 (A^{(1)}) 
\eeq
where $C^{(1)}$ is the dynamical $\IZ_n$  gauge field of an $\IZ_n$ TQFT and $\omega_3 (A^{(1)})$ is 3d $U(1)$ Chern-Simons action as defined in eq.~(\ref{eq:omega_3_3d_CS}). 

The implications of this TQFT-coupling via discrete gauging can be summarized as follows.
\begin{itemize}
\item $\left[ \right.$Section~\ref{subsubsec:discrete gauging_free GB}$\left. \right]$ In order to identify the important physical features of discrete gauging, we first discuss discrete gauging in \emph{free} $U(1)$ Goldstone boson theory, i.e.~$K_A=0$ case. We show two consequences of discrete gauging of shift symmetry of Goldstone boson $\phi (x) \to \phi (x) + c$: (i) it projects out some of the local operators $I(q,x) = e^{iq \phi (x)}$ and (ii) it adds additional cosmic strings with \emph{fractional} winding numbers. Specifically, if we gauge a $\IZ_n$ subgroup of the axion shift symmetry, local operators with $q \neq n \mathbb{Z}$ are removed since they are not invariant under $\IZ_n$ gauge transformations. Simultaneously, surface operators (i.e.~cosmic strings) with fractional winding
\beq
\oint \frac{d \phi}{2\pi} \in \frac{1}{n} \mathbb{Z}
\eeq
are included; these objects are identified as cosmic strings of the TQFT sector.
\item $\left[ \right.$Section~\ref{subsubsec:discrete gauging Axion-Maxwell}$\left. \right]$ Next, we describe axion-Maxwell theory coupled to a TQFT via discrete gauging of a $\IZ_n$ subgroup of the axion shift symmetry. Here, we show how the spectra of local operator and cosmic string are modified in the presence of TQFT-coupling.
\item $\left[ \right.$Section~\ref{subsec:UV theory_TQFT coupling II}$\left. \right]$ We present a KSVZ-type UV field theory which flows in the IR limit to the axion-Maxwell theory coupled to a $\IZ_n$ TQFT as described above. This result further illustrates our claim that non-trivial TQFT-couplings can appear from standard UV field theories. 
\item $\left[ \right.$Section~\ref{subsec:3-group_discrete_gauging}$\left. \right]$ We show how $\IZ_n$ discrete gauging changes the 3-group global symmetry structure of the theory. In particular, in addition to above described features, the 1-form electric symmetry is broken $\mathbb{Z}_K^{(1)} \to \mathbb{Z}_{K/n}^{(1)}$, thus altering the spectrum of Wilson lines. 
\item $\left[ \right.$Section~\ref{subsec:other TQFT couplings}$\left. \right]$  It is possible to systematically classify the possible ways to couple axion-Maxwell theory to TQFTs  via discrete gauging. We list possible discrete gaugings of the axion-Maxwell theory and briefly discuss the resulting symmetry structure for each case. 
\end{itemize}

In summary, the analysis in this paper demonstrates the importance of understanding couplings to TQFTs and tracking them along RG flows in particle physics settings. Further, the fact that the models with TQFT-couplings we consider in this paper have such simple UV completions illustrates that such couplings  are not at all exotic or unrealistic. Instead, they generically arise in the long distance behavior of standard QFTs. We hope that our work demonstrates the utility of the techniques provided by generalized global symmetries in a concrete setup and helps to initiate a broader effort in the application of generalized global symmetry techniques in particle physics.

\section{TQFT-Coupling I: Axion-Portal to a TQFT}
\label{sec:TQFT-coupling I}

In this section, we discuss an axion-Maxwell theory coupled to a $\mathbb{Z}_n$ topological quantum field theory. The action is presented in eqns.~(\ref{eq:intro_aMW})~--~(\ref{eq:intro_couplingI}). For convenience of the discussion, we reproduce them below. 
The axion-Maxwell sector without a TQFT-coupling is described by the action
\beq
S_0 = \frac{1}{2} \int  d a \wedge * d a + \frac{1}{2 g^2} \int F_A^{(2)} \wedge * F_A^{(2)} - \frac{iK_{A}}{8\pi^2 f_a} \int a F_A^{(2)} \wedge F_A^{(2)}~.
\label{eq:S_0}
\eeq
Here, $a$ is a periodic scalar field (axion) with $a \sim a + 2\pi f_a$ (which is the remnant of a $U(1)_{\rm PQ}$ global symmetry that is spontaneously broken at a scale set by the axion decay constant $f_a$) and $A^{(1)}$ and $F_A^{(2)}$ are $U(1)_A$ one-form gauge field and its two-form field strength, respectively. The coupling constant $K_A\in \IZ$ is quantized, which is necessary in order to ensure the 
periodicity of $a$, and furthermore  is the coefficient of a perturbative $\left[ U(1)_{\rm PQ} U(1)_A^2 \right]$ Adler-Bell-Jackiw (ABJ) anomaly \cite{Adler:1969gk, Bell:1969ts}. 

We would now like to study what happens when we couple axion-Maxwell sector to a TQFT. One can imagine a various of ways to achieve this. 
A simple choice of TQFT to couple to is a $\mathbb{Z}_n$ gauge theory \cite{Horowitz:1989ng, Horowitz:1989km, Banks:2010zn, Kapustin:2014gua, Gaiotto:2014kfa} which can be viewed as the low energy limit of a spontaneously broken $U(1)_B$ gauge theory. We can couple such a  $\mathbb{Z}_n$ TQFT to the axion-Maxwell theory via an axion portal coupling\footnote{The factor of 2 difference between $K_{AB}$ and $K_B$ (and $K_A$) terms comes from the fact that the anomaly polynomial is given by $ch_2 (F^{(2)})=\Tr[e^{\frac{i F^{(2)}}{2\pi}}]$ restricted to the degree 4 differential form where the trace is over the total bundle. The cross term comes with a factor of 2 due to the fact that a $U(1)_A\times U(1)_B$ bundle has $ch_2(F_A+F_B)=\frac{1}{2!\times (2\pi^2)} \left(F_A^{(2)}+F_B^{(2)} \right)^2$. }
\bea
S_1 = \frac{i n}{2\pi} \int B^{(2)} \wedge F_B^{(2)} - \frac{i K_{AB}}{4 \pi^2 f_a} \int a F_A^{(2)} \wedge F_B^{(2)} - \frac{i K_{B}}{8\pi^2 f_a} \int a F_B^{(2)} \wedge F_B^{(2)}
\label{eq:S_1}
\eea
where $F_B^{(2)} = d B^{(1)}$ is the field strength of a one-form $\mathbb{Z}_n$ gauge field $B^{(1)}$ and $B^{(2)}$ is a two-form gauge field associated with one-form $\mathbb{Z}_n^{(1)}$ gauge invariance.\footnote{In fact, it is possible and interesting to consider coupling to a broader class of $\IZ_n$ TQFTs such as \cite{Kapustin:2014gua} 
\beq
S_{{\rm BF}'} = \frac{in}{2\pi} \int B^{(2)} \wedge F_B^{(2)} + \frac{inp}{4\pi} B^{(2)} \wedge B^{(2)}.
\eeq
}
This theory also admits additional interaction terms such as the topological $\theta$-term $S=\frac{i\theta}{2\pi}\int da\wedge dB^{(2)}$ or we may add a ``kinetic mixing'' $F_A^{(2)} \wedge * F_B^{(2)}$ and study its effects. In this paper, we do not consider these additional terms and leave their interesting possibility for future investigations. 

The first term in eq.~(\ref{eq:S_1}) is the action for $\mathbb{Z}_n$ TQFT, often also called a BF theory. It describes a $\mathbb{Z}_n$ gauge theory and a brief review is presented in Appendix~\ref{app:review_BF} (see \cite{Horowitz:1989ng, Horowitz:1989km, Banks:2010zn, Kapustin:2014gua, Gaiotto:2014kfa} for more discussion). The BF theory sector admits two gauge invariant ``electric'' operators, a Wilson line and a Wilson surface operators: %
\beq
W_1 (\Sigma_1, \ell) = e^{i \ell \oint_{\Sigma_1} B^{(1)}}~,\quad W_2 (\Sigma_2, m) = e^{i m \oint_{\Sigma_2} B^{(2)}}
\label{eq:BF_line_surface_operators}
\eeq
which act as sources for $B^{(2)}$ and $B^{(1)}$ respectively due to form of the equations of motion.

The second and third terms in eq.~(\ref{eq:S_1}) describe (local) interactions between the axion-Maxwell sector and the TQFT sector. 
The goal of the rest of this section is to investigate the implications of this TQFT-coupling. In particular, we are interested in properties of the IR effective theory described by $S_0 + S_1$ which are \emph{universal}, i.e.~independent of specific UV completions. 

In the rest of this section we study the implication of the TQFT on the 2d QFT living on the axion-string and how this physics can be realized in a simple UV model. 
The coupling discussed in this section has the possibility to produce interesting, observational signals. We relegate these details to Section~\ref{sec:pheno} where we more broadly discuss the phenomenological implications of coupling axion-Maxwell theory to TQFTs.

\subsection{Anomaly Inflow and 2d String Worldsheet QFT} 
\label{subsec:anomaly inflow}

In this section, we discuss the axion strings in the theory described by $S_0 + S_1$ shown in eq.~(\ref{eq:S_0}) and (\ref{eq:S_1}), with a special emphasis on the universal IR  features. 

Imagine a cosmic string placed in the spacetime $M_4$.  We would like to study the effects on the world volume induced by the axionic couplings
\beq
S_{\rm axion} = - \frac{iK_A}{8\pi^2 f_a} \int a F_A^{(2)} \wedge F_A^{(2)} - \frac{i K_{AB}}{4\pi^2 f_a} \int a F_A^{(2)} \wedge F_B^{(2)} - \frac{i K_{B}}{8\pi^2 f_a} \int a F_B^{(2)} \wedge F_B^{(2)}~.
\label{eq:S_anom1}
\eeq
To simplify expressions, we often use the language of descent equation of chiral anomaly (see for e.g.~\cite{Harvey:2005it, Weinberg:1996kr, Hong:2020bvq})
\eq{
 \omega_4 = &\frac{1}{8\pi^2} F^{(2)} \wedge F^{(2)} = d \omega_3~,\quad 
\omega_3 = \frac{1}{8\pi^2} A^{(1)} \wedge F^{(2)}~, \label{eq:omega_3_3d_CS}\\
 \delta_\alpha \omega_3 =& \,d \omega_2~,~~~\hspace{3cm}  \omega_2 = \frac{\alpha}{8\pi^2} F^{(2)}~.
}
We comment that the 3d Chern-Simons (CS) action $\omega_3$ 
is defined by a cohomology class (shifting by an exact term corresponds to shifting by local counterterms) and expressions shown are canonical choices (see appendices of \cite{Hong:2020bvq} for a comprehensive review). 
This allows us to rewrite $S_{\rm axion}$ as 
\bea
S_{\rm axion} &=& - \frac{i}{f_a} \int a \left( K_A \, \omega_4 (A^{(1)}) + 2 K_{AB} \, \omega_4 (A^{(1)},B^{(1)}) + K_B \, \omega_4 (B^{(1)}) \right) \nonumber \\
&=&  \frac{i}{f_a} \int da \wedge \left( K_A \, \omega_3 (A^{(1)}) + 2 K_{AB} \, \omega_3 (A^{(1)},B^{(1)}) + K_B \, \omega_3 (B^{(1)}) \right)~.
\eea
In the presence of the axion vortex, neither $a$ nor $da$ is well-defined at the vortex worldsheet $r=0$.\footnote{By this, we really mean that 
at distance scale smaller than $f_a$, the axion winding is not anymore protected by topology, i.e.~there is enough energy fluctuation to unwind the axion. In the deep IR, the region $r < f_a^{-1}$ is represented as a singularity around which axion winds, but in the UV it is a smooth field configuration.  } However, as demonstrated in \cite{Harvey:2000yg} we can obtain an action that is well-defined everywhere in $M_4$ even in the presence of an axion string. This can be achieved by ``smoothing out'' the axion string singularity by inserting a bump function $\rho (r)$ into $S_{\rm axion}$:
\beq
S_{\rm axion} =  \frac{i}{f_a} \int (1+\rho) \, da \wedge \left( K_A \, \omega_3 (A^{(1)}) + 2 K_{AB} \, \omega_3 (A^{(1)},B^{(1)}) + K_B \, \omega_3 (B^{(1)}) \right)~.
\label{eq:S_anom2}
\eeq
so that $(1+\rho(r))\to 0$ smoothly as $r\to 0$ and $\rho(r)=0$ for $r>1/f_a$ . 
This insertion of $1+\rho(r)$ regularizes the the action 
in the string core $r=0$, extending the validity of our description there. 
In addition, we recover the original action well outside the vortex due to the behavior of $\rho(r)$ for $r>1/f_a$. As we discuss below, another advantage of implementing the  bump function regulator  is that it 
allows us to switch from covariant to consistent anomalies on the axion string world sheet so that the anomaly cancellation 
 can be understood straightforwardly \cite{Naculich:1987ci,Harvey:2000yg}. 
In summary, $\rho (r)$ can be thought of as  a smooth generalization of a $\delta$-function that describes the embedding of the cosmic string world-sheet $M_2^{\rm st}$ into $M_4$. 

With these conditions on $\rho(r)$, a cosmic string with winding number $m$ satisfies
\beq
m = \oint_{S^1} (1+\rho) \frac{da}{2\pi f_a} = \int_{D_2} d \left[ (1+\rho) \frac{da}{2\pi f_a} \right] \;\;\; \Longrightarrow  \;\;\; d \rho \wedge d a = 2\pi m f_a \delta^{(2)} (M_2^{\rm st})~.
\label{eq:rho_a_delta}
\eeq
Here, $\delta^{(2)} (M_2^{\rm st})$ is the two-form $\delta$-function, that is only non-vanishing only on the $(1+1)$d cosmic string world-sheet $M_2^{\rm st}$.

\subsubsection{Anomaly Inflow}
\label{subsubsec:anomaly_inflow}

Now let us consider the axion interaction term $S_{\rm axion}$. Under a $U(1)_A$ gauge transformation, $A^{(1)} \to A^{(1)} + d \lambda_A$, the action varies as 
\bea
\delta_A S_{\rm axion} &=& \frac{i}{8\pi^2 f_a} \int (1+\rho) da \wedge \left( K_A d \lambda_A \wedge F_A^{(2)} + 2 K_{AB} d \lambda_A \wedge F_B^{(2)} \right)  \nonumber \\
&=& \frac{im}{4\pi} \int_{M_4}  \lambda_A \delta^{(2)} (M_2^{\rm st}) \wedge \left( K_A F_A^{(2)} + 2 K_{AB} F_B^{(2)} \right)~.
\label{eq:inflow_A}
\eea
Here we see that the variation leads to an anomalous term that is localized on the cosmic string worldsheet. Similarly, the variation of $S_{\rm axion}$ under $\IZ_n$ gauge transformations: $B^{(1)} \to B^{(1)} + d \lambda_B$, 
is given by 
\beq
\delta_B S_{\rm axion} = \frac{im}{4\pi} \int_{M_4}  \lambda_B \delta^{(2)} (M_2^{\rm st}) \wedge \left( 2 K_{AB} F_A^{(2)} + K_{B} F_B^{(2)} \right)~.
\label{eq:inflow_B}
\eeq
Since our theory is well defined everywhere, 
the anomalies localized on the cosmic string world-sheet in eqns.~(\ref{eq:inflow_A}) and ({\ref{eq:inflow_B}}) must be canceled which furthermore implies that there must be degrees of freedom living on the cosmic string worldsheet that cancel these anomalies.  In terms of 2d dynamics, these anomalies must be reproduced by \emph{consistent} gauge anomalies \cite{Naculich:1987ci,Harvey:2000yg}: 
a $U(1)_A^2$ gauge anomaly with coefficient $-K_A$, a $U(1)_A\times \mathbb{Z}_n$, mixed anomaly with 
coefficient $-K_{AB}$, and a $\IZ_n^2$ gauge anomaly with coefficient $-K_B$.

It is also illuminating to reproduce our discussion on the existence of charged matter on the cosmic string world volume in terms of currents. The current is defined by taking a functional derivative with respect to $A^{(1)}$,
\eq{
* J_1 (a, A^{(1)}) =& K_A \left( \frac{1}{4\pi^2 f_a} (1+\rho) d a \wedge F_A^{(2)} - \frac{1}{8\pi^2 f_a} d \rho \wedge d a \wedge A^{(1)} \right) \\
&+ \frac{K_{AB}}{4\pi^2 f_a} (1+\rho) d a \wedge F_B^{(2)}
\label{eq:J_A}
}   
which satisfies the conservation equation

\beq
d * J_1 (a, A^{(1)}) = \frac{K_A}{4\pi} \delta^{(2)} (M_2^{\rm st}) \wedge F_A^{(2)} + \frac{K_{AB}}{2\pi} \delta^{(2)} (M_2^{\rm st}) \wedge F_B^{(2)}
\label{eq:d J_A}
\eeq
in the presence of an axion string with unit winding. This shows that there are indeed charged matter fields that are localized on the world sheet of the cosmic string. 

Here there is a subtlety to the matching of the gauge anomalies involving the $\IZ_n$ gauge symmetry. The point is that at short distances, the $\IZ_n$ gauge symmetry can be enhanced to a $U(1)_B$ gauge symmetry. 
In these cases, the $\IZ_n$ anomaly cancellation is only required 
${\text{mod}} \; n$. 
We can see this reduction of the anomaly coefficients mod $n$ as follows. First note that any terms of the form
\beq
S_{\rm ct} = \frac{i n}{8\pi^2 f_a} \int (1+\rho) da \wedge \left( 2 z_1 B^{(1)} \wedge F_A^{(2)} + z_2 B^{(1)} \wedge F_B^{(2)}  \right)
\label{eq:S_ct}
\eeq
with $z_{1,2} \in \mathbb{Z}$ are gauge invariant. This term is clearly  $U(1)_A$  invariant, so we only need to discuss $\mathbb{Z}_n$ invariance. Under $\mathbb{Z}_n$ gauge transformations, 
the above counterterms transform as
\beq
\delta  S_{\rm ct}  = 2\pi i k\int \frac{d \rho \wedge da}{2\pi f_a} \wedge \left( z_1 \frac{F_A^{(2)}}{2\pi} + z_2 \frac{F_B^{(2)}}{2\pi} \right) ~.
\eeq
Comparing with eq.~(\ref{eq:S_anom1}) and (\ref{eq:S_ct}), we see that the local counterterms eq. \eqref{eq:S_ct} allow us to reduce the anomaly coefficients for $\IZ_n$ ${\rm mod} \; n$,  which of course is what we expect for anomalies of $\IZ_n$ gauge groups \cite{Csaki:1997aw}.

\subsubsection{Anomaly Cancellation by Fermion Zero Modes}
\label{subsubsec:fermion_zero_mode_IR}

As we have shown, the consistency of IR EFT requires that anomalies in eq.~(\ref{eq:inflow_A}) and (\ref{eq:inflow_B}) need to be canceled by a 2d QFT on the cosmic string world sheet. These anomalies can be matched by 2d massless chiral fermions that are localized on the string. And indeed, such 2d fields often arise in UV completions as perturbations around zero-modes of bulk 4d fermions in the presence of cosmic strings 
\cite{Jackiw:1981ee}.  
We will demonstrate this feature of anomaly matching in a particular UV completion in Section  \ref{subsec:UV model}. 

Here, we will argue for the existence of 2d chiral fermions on the cosmic string world sheet purely based on IR consistency and determine conditions for their quantum numbers. For a set of $(1+1)$d Weyl fermions $\{ \alpha_i (z, t) \}$ living on the core of the cosmic string, if their charges under $U(1)_A \times \mathbb{Z}_n$ are $(Q_i, k_i)$, the anomaly cancellation conditions are written as
\bea
\sum_i Q_i^2 = - K_A~,\quad 
\sum_i Q_i k_i = - K_{AB} \; {\rm mod} \; n~,\quad 
\sum_i k_i^2 = - K_B \; {\rm mod} \; n~.
\label{eq:conditions}
\eea
This is a straightforward but interesting result. 
Adding the TQFT-coupling, while not altering the 4d bulk QFT, has modified the 2d QFT on the string worldsheet: anomaly cancellation requires that the chiral degrees of freedom carry $\mathbb{Z}_n$ as well as $U(1)_A$ charges. In particular, as shown by the second requirement in eq.~(\ref{eq:conditions}), at least some of the zero mode fermions localized on the string must be charged under both $U(1)_A$ and $U(1)_B$.  As we discuss in Section~\ref{sec:pheno}, this property can lead to interesting features in cosmic string physics, which could in principle be observable.

\subsection{UV Field Theory Completion}
\label{subsec:UV model}

In this section, we introduce a UV completion that reduces in the IR to the effective theory described by eq.~(\ref{eq:S_0}) and (\ref{eq:S_1}). We solve the Dirac equation in the vortex background and determine the spectrum of fermion zero modes. Using this, we will check the anomaly cancellation discussed in Section~\ref{subsubsec:fermion_zero_mode_IR} explicitly.

\subsubsection{UV theory}

Here we take a $U(1)_A\times U(1)_B$ gauge theory coupled to scalars and fermions. The Lagrangian is 
\bea
\mathcal{L} = && - \frac{1}{4 g_A^2} F_A^{(2)} \wedge * F_A^{(2)} - \frac{1}{4 g_B^2} F_B^{(2)} \wedge * F_B^{(2)} +  \sum_{i=1}^2 \bar{\psi}_i i \slashed{D} \psi_i + \bar{\chi}_i \slashed{D} \chi_i \nonumber  \\
&& - |D \Phi_1|^2 - |D \Phi_2|^2 -  \lambda_1 \Phi_1^\dagger \psi_1 \chi_1 - \lambda_2 \Phi_2 \psi_2 \chi_2 + {\rm h.c.} + V(\Phi_1, \Phi_2)~.
\label{eq:UV_theory_1}
\eea
The scalar potential is such that both $\langle \Phi_1 \rangle = f_1$ and $\langle \Phi_2 \rangle = f_2$ are non-zero. 
The quantum numbers of the fields are summarized in Table~\ref{tab:charges1}. 

The vacuum expectation value (vev) $f_2$ spontaneously breaks $U(1)_B \to \mathbb{Z}_n$, producing a $\IZ_n$ BF theory, while $f_1$ spontaneously breaks a combination of global $U(1)_{\rm PQ}$ and $U(1)_B$, which leads to a low energy axion. The fermion covariant derivatives contain $U(1)_A$ and $U(1)_B$ gauge fields with their corresponding charges. 

\begin{table}
\center
\begin{tabular}{|c||c|c|c|}
\hline
  & $U(1)_{\rm PQ}$ & $U(1)_A$ &  $U(1)_B$ \\
\hline\hline
$\Phi_1$ & $1$ & $0$ & $n$ \\
$\Phi_2$ & $0$ & $0$ & $n$ \\
\hline
$\psi_1$ & $1$ & $1$ & $q$ \\
$\chi_1$ & $0$ & $-1$ & $n-q$ \\
\hline
$\psi_2$ & $0$ & $1$ & $q-n$ \\
$\chi_2$ & $0$ & $-1$ & $-q$ \\
\hline
\end{tabular}
\caption{Quantum numbers of the fields of the UV completion eq.~(\ref{eq:UV_theory_1}). This theory admits the limit $\Lambda \gg f_a$ but not $f_a \gg \Lambda$.}
\label{tab:charges1}
\end{table}
To get an effective theory below $f_1$ and $f_2$ by integrating out heavy fermions, it is convenient to rotate away the phases of scalar fields from the Yukawa terms. We first write the scalar fields as $\Phi_i = \varphi_i e^{i \theta_i}, i = 1,2$.  
For vacuum solution $\varphi_i = f_i$, while for a string solution (with winding number $n_i$) the Higgs vev has a non-trivial profile 
\beq
\varphi_i = F_i (r)
\eeq
while $\theta_i$ winds around the core of the string. 
The phases $\theta_i$ are (would-be) Goldstone bosons and can be removed from the Yukawa terms by the following field redefinitions:
\beq
\psi_1 = e^{i \theta_1} \hat{\psi}_i~,\quad 
\chi_2 = e^{-i \theta_2} \hat{\chi}_2
\label{eq:field_redef}
\eeq 
which are associated to the global symmetry transformations $U(1)_1$ and $U(1)_2$ respectively. 
These are, however, anomalous transformations and which will generate axionic terms in the effective action.
\beq
S_{\rm axion} = \frac{i}{8\pi^2} \int (\theta_1 - \theta_2 ) \left( F_A^{(2)} \wedge F_A^{(2)} + 2 q F_A^{(2)} \wedge F_B^{(2)} + q^2 F_B^{(2)} \wedge F_B^{(2)} \right)
\label{eq:S_anom_in_UV_theory}
\eeq
where the prefactors $1, q$ and $q^2$ are respectively anomaly coefficients for $U(1)_i\times U(1)_A^2$, $U(1)_i \times U(1)_A \times U(1)_B$, and $U(1)_i\times  U(1)_B^2$ anomalies ($i=1,2$).
After these field redefinitions, the Yukawa interactions become just fermion mass terms at energies below $f_1$ and $f_2$. At these energies, the effective action is eq.~(\ref{eq:S_0}) and (\ref{eq:S_1}) with the matching
\bea
&& \theta_1 - \theta_2 = \frac{\Pi_1}{f_1} - \frac{\Pi_2}{f_2} \equiv \frac{a}{f_a}~,\quad f_a = \frac{f_1 f_2}{\sqrt{f_1^2 + f_2^2}}~,\label{eq:thetas_axion} \\
&& K_A = 1~,\quad K_{AB} = q~,\quad K_B = q^2~.
\eea
The other orthogonal combination of $\theta$'s is the would-be Goldstone boson eaten by the $U(1)_B$ gauge boson.
In the IR  the eaten Goldstone boson leads to a $\IZ_n$ gauge theory since it has charge $n>1$. See Appendix~\ref{app:review_BF} for more details.

In the limit $f_2 \gg f_1$, this theory describes the case where $U(1)_B$ is Higgsed down to $\mathbb{Z}_n$ gauge theory below $f_2$ generating BF strings. 
Provided that $\lambda_2$ is not too small, the pair $\psi_2$ and $\chi_2$ are integrated out around $f_2$, and the effective theory at $f_2\gg E\gg f_1$ is a KSVZ-type theory \cite{Kim:1979if, Shifman:1979if} coupled to a $\mathbb{Z}_n$ BF theory. This latter coupling is in the form of anomalous terms proportional to $\theta_2$ in eq.~(\ref{eq:S_anom_in_UV_theory}). Here, $\theta_2$ encodes the 2-form BF degree of freedom $B^{(2)}$ via the 4d duality relation $d \theta_2 \sim * d B^{(2)}$ so that the ``BF''-sector of the theory is described by 
\beq
\mathcal{L} = \frac{in}{2\pi} B^{(2)} \wedge F_B^{(2)} + \frac{i}{8\pi^2} q^2 * d B^{(2)} \wedge B^{(1)} \wedge F_B^{(2)}~.
\eeq
At lower energies with $E\ll f_1$,  the $U(1)_{\rm PQ}$ is broken by the vev of $\Phi_1$, resulting in the physical axion field in $S_{\rm axion}$ (see eq.~(\ref{eq:thetas_axion})).

In cosmology, at temperature $f_1 < T < f_2$, a $\mathbb{Z}_n$ BF theory arises and BF strings can form according to the Kibble-Zurek mechanism \cite{Kibble:1980mv, Kibble:1976sj, Zurek:1996sj, Zurek:1985qw}. The BF string is supported by non-zero $B^{(1)}$ magnetic flux through its core and causes $\theta_2$ to wind around it in the dual picture. Equivalently, the BF string defect can be thought of as the Wilson surface operator for $B^{(2)}$ as defined in eq.~(\ref{eq:BF_line_surface_operators}).  

We make a few remarks about the Wilson operators of the BF theory. BF theory has $\IZ_n$-classified line operators $W_1 = e^{i \int_{\Sigma_1} B^{(1)}}$ that are charged under a 1-form $\mathbb{Z}_n^{(1)}$ global symmetry which acts as $B^{(1)} \to B^{(1)} + \frac{2\pi}{n} \lambda^{(1)}$, where $\lambda^{(1)}$ is a 1-form with integer periods.
These operators can be cut by $\IZ_n$ charged fermions -- in our model ~$\{ \psi_1, \chi_1 \}$. Therefore, the topologically protected line operators are characterized by $\mathbb{Z}_{{\rm GCD}(q,n-q)}$.

Additionally, the theory has $\mathbb{Z}_n$-classified string/surface operators $W_2 = e^{i \int_{\Sigma_2} B^{(2)}}$ that are charged under a 2-form $\mathbb{Z}_n^{(2)}$ global symmetry which acts as $B^{(2)} \to B^{(2)} + \frac{2\pi}{n} \lambda^{(2)}$ where $\lambda^{(2)}$ is a 2-form with integer periods. They may be cut and broken by creation of monopole-anti-monopole pair. In the absence of monopoles in the theory, all $\mathbb{Z}_n$ BF strings are topologically stable.\footnote{$U(1)$ gauge theory with only electrically charged particles has one-form magnetic \emph{global} symmetry characterized by $d F =0$. It is believed that quantum gravity does not admit any exact global symmetry and so this symmetry should be either gauged or explicitly broken. If $U(1)$ is embedded in a GUT group, then monopoles exist and they explicitly break the 1-form magnetic symmetry.}

These $\mathbb{Z}_n$ BF strings are quasi-Aharonov-Bohm strings according to the classification by Polchinski \cite{Polchinski:2005bg, Banks:2010zn}. Indeed, the light fermions at $f_1 < E < f_2$ give rise to discrete Aharonov-Bohm phases when they circle around the BF strings. These strings are ``quasi'' because the probe states are not only charged under $\mathbb{Z}_n\subset U(1)_B$, but they are also charged under low energy unbroken gauge group $U(1)_A$.

At energies below $f_1$, all charged fermions are integrated out.
Since there are no charged states, at this scale BF strings become \emph{local strings} which do not have any topological charges probable by observers at spatial infinity. In addition, due to the $U(1)_{\rm PQ}$ breaking, we get \emph{global strings} which is measured by a topological charge: the axion winding.\footnote{In a theory of quantum gravity, the global shift symmetry of the axion is likely to be broken explicitly by instanton effects (due to the expectation that there are no global symmetries in quantum gravity) which generates an axion potential. This spontaneously breaks the discrete shift symmetry and leads to formation of domain walls bounded by the axion string. This then destabilizes the axion strings because such a domain wall will tend to shrink the string or a hole of closed string can be nucleated inside of the wall.}

Our theory with charge assignment given in Table~\ref{tab:charges1}, however, does not allow for the $\mathbb{Z}_n$ TQFT to emerge below PQ symmetry breaking. Note that as we try to take the other limit $f_1 \gg f_2$, $U(1)_B$ is still broken to $\mathbb{Z}_n$ since both scalars carry $U(1)_B$ charge $n$.  However, it is possible to construct models which have $f_a\gg \Lambda_{\rm BF}$.

\subsubsection{Fermion Zero Modes} 
 
In this section, we solve the Dirac equation of the UV theory described by the action in eq.~(\ref{eq:UV_theory_1}) with the charge assignment in Table~\ref{tab:charges1}, and determine the spectrum of fermion zero modes, thereby demonstrating anomaly cancellation in the presence of an axion string. Our analysis below closely follows \cite{Witten:1984eb}.

The equations of motion for the pair $\{ \psi_1, \chi_1 \}$ are given by
\bea
i \slashed{D} \psi_1 = \lambda_1 \Phi_1 \overline{\chi}_1~,\qquad  i \slashed{D} \chi_1 = \lambda_1 \Phi_1 \overline{\psi}_1
\eea
We look for a solution in the background of string $\Phi_1 = F_1 (r) e^{i \phi}$ of the form\footnote{The case with arbitrary winding number can be studied by adopting the method of \cite{Jackiw:1981ee}. }
\beq
\psi_1 (x, y, z, t) = \alpha_1 (z, t) \beta_1(x, y)~,\qquad \chi_1 (x, y ,z, t) = \eta_1 (z, t) \xi_1 (x, y)
\eeq
where $\alpha_1$ and $\eta_1$ are zero modes localized on the string core and $\beta_1$ and $\xi_1$ are transverse zero modes. 
We split the Dirac operator into string- and transverse-part $i \slashed{D} = i \slashed{D}_s (z,t) + i \slashed{D}_T (x,y)$ and we also define chirality operators $\Gamma^{\rm int} = \gamma^0 \gamma^3$ and $\Gamma^{\rm ext} = i \gamma^1 \gamma^2$ that act on the string-core and transverse space, respectively. 
 
Let us first solve for the transverse zero modes, which satisfy
\beq
i \slashed{D}_T \beta_1 = \lambda_1 \Phi_1 \overline{\xi}_1~,\qquad i \slashed{D}_T \xi_1 = \lambda_1 \Phi_1 \overline{\beta}_1.
\eeq
For $\phi$-independent solution, the transverse Dirac operator can be written as
\beq
i \slashed{D}_T = i \left( \gamma^1 \partial_1 + \gamma^2 \partial_2 \right) = i \gamma^1 \left( \cos \phi + i \Gamma^{\rm ext} \sin \phi \right) \partial_r
\eeq
and the Dirac equations in the unit winding string background become 
\bea
&& i \gamma^1 \left( \cos \phi + i \Gamma^{\rm ext} \sin \phi \right) \partial_r \beta_1 = \lambda_1 F_1 e^{i \phi} \overline{\xi}_1 ~, \label{eq:Dirac_eq_1} \\
&& i \gamma^1 \left( \cos \phi + i \Gamma^{\rm ext} \sin \phi \right) \partial_r \xi_1 = \lambda_1 F_1 e^{i \phi} \overline{\beta}_1~. \label{eq:Dirac_eq_2}
\eea
The angular dependence requires that
\beq
\Gamma^{\rm ext} \beta_1 = + \beta_1~,\qquad \Gamma^{\rm ext} \xi_1 = + \xi_1
\label{eq:transverse_zero_mode_chirality_1}
\eeq
and with this, the radial part can be straightforwardly integrated to obtain
\beq
\beta_1 (r) = {\rm exp} \left( - \int_0^r \lambda_1 F_1 (r' ) dr' \right)~,\qquad  \overline{\xi}_1 = - i \gamma^1 \beta_1.
\eeq

Using the form of the transverse zero modes, the equation for the string zero modes reduces to 
\beq
\left( i \slashed{D}_s \alpha_1 \right) \beta_1 = \left( \overline{\eta}_1 - \alpha_1 \right) \left( \lambda_1 \Phi_1 \overline{\xi}_1 \right)~. 
\eeq
This shows that non-trivial string zero mode exists if and only if $\alpha_1 = \overline{\eta}_1$ and that is there is only one string zero mode per pair $\{ \psi_1, \chi_1 \}$. The string zero mode equation becomes
\bea
0 = i \slashed{D}_s \alpha_1 = i \gamma^0 \left( \partial_t + \Gamma^{\rm int} \partial_z \right) \alpha_1 (z,t)~.
\eea
Recalling $\gamma^5 \psi_1 = - \psi_1$, eq.~(\ref{eq:transverse_zero_mode_chirality_1}) implies that $\alpha_1$ should be 2d LH chiral fermion, $\Gamma^{\rm in} \alpha_1 = - \alpha_1$. The zero mode equation is then solved by
\beq
\alpha_1 (z,t) = g (z+t)
\eeq
for an arbitrary function $g$. This means that 2d left-handed chiral zero mode propagates in the $-\hat{z}$ direction at the speed of light.
Overall, in the background of $\Phi_1$-string of winding number one, the pair $\{ \psi_1, \chi_1 \}$ coupled to $\Phi_1$ gives rise to a single left-handed 2d zero mode traveling in the $-\hat{z}$ direction at the speed of light. Analogously, on a anti-$\Phi_1$ string, the same procedure shows that $\beta_1$ should be a negative helicity state while $\alpha_1$ is now a positive helicity mode running in the $+ \hat{z}$ direction.

The analysis for $\{ \psi_2, \chi_2 \}$ coupled to $\Phi_2$ follows similarly with one exception: while $\{ \psi_1, \chi_1 \}$ couples to $\Phi_1^\dagger$, $\{ \psi_2, \chi_2 \}$ couples to $\Phi_2$ (see eq.~(\ref{eq:UV_theory_1})). This has an effect that the pair $\{ \psi_2, \chi_2 \}$ coupled to winding number one $\Phi_2$-string behaves like they couple to anti-$\Phi_2$-string (winding $-1$). Practically, we only need to replace $e^{i \phi} \to e^{-i \phi}$ in eq.~(\ref{eq:Dirac_eq_1}) and (\ref{eq:Dirac_eq_2}) together with relabeling $1 \to 2$. The final result is the pair $\{ \psi_2, \chi_2 \}$ in the background of winding number one $\Phi_2$-string has a single RH zero mode localized at the core which propagates in the $+ \hat{z}$ direction.

The quantum numbers of string zero modes are read off from those of original fermions after taking into account the field redefinitions eq.~(\ref{eq:field_redef}) (note that the Dirac equations are in the basis where all the phase degrees of freedoms are removed from the Yukawa interaction).\footnote{Additionally, we must take into account the change of charges due to the field redefinitions so that the condition $\alpha_1 = \overline{\eta}_1$ can be fulfilled. } The final results are summarized in Table~\ref{tab:charges3}.
\begin{table}
\center
\begin{tabular}{|c||c|c|c|c|}
\hline
  & chirality & $U(1)_{\rm PQ}$ & $U(1)_A$ &  $U(1)_B$ \\
\hline\hline
$\alpha_1$ & LH & $0$ & $1$ & $q-n$ \\
\hline
$\alpha_2$ & RH & $0$ & $1$ & $q-n$ \\
\hline
\end{tabular}
\caption{Quantum numbers of string zero modes in the $\Phi_1$-string of winding number one (for $\alpha_1$) and the $\Phi_2$-string of winding number one (for $\alpha_2$).}
\label{tab:charges3}
\end{table} 

\subsubsection{Anomaly Cancellation}

Recall from eq.~(\ref{eq:S_anom_in_UV_theory}) that at $E < f_1, f_2$, the phase factor appearing in the anomalous term is the combination $\theta_1 - \theta_2 = a / f_a$. Here, $\theta_1$ measures the winding of $\Phi_1$-string and $\theta_2$ measures that of $\Phi_2$-string. Let us denote the winding numbers as $\{n_1, n_2\}$. We are interested in checking the anomaly cancellation in the background of a string with winding number $\{n_1, n_2\}$. In this case, the regularizing string satisfies
\beq
d \rho \wedge d a = 2\pi (n_1 - n_2) f_a\, \delta^{(2)} \left( M_2^{\rm st} \right)~.
\eeq
Using this, anomalous variation of $S_{\rm axion}$ is found to be 
\bea
&& \delta_A S_{\rm axion} =  \frac{1}{4\pi} \int_{M_2^{\rm st}} (n_1 - n_2) \lambda_A \left( F_A^{(2)} + q F_B^{(2)} \right)~, \label{eq:UV_theory_bulk_anomaly_1}\\
&& \delta_B S_{\rm axion} =  \frac{1}{4\pi} \int_{M_2^{\rm st}} (n_1 - n_2) \lambda_B \left( q F_A^{(2)} + q^2 F_B^{(2)} \right)~. \label{eq:UV_theory_bulk_anomaly_2}
\eea
One immediately notices that for $n_1 = n_2$, there is no localized anomaly on the composite cosmic string. The reason is that the 2d fermions $\alpha_1$ and $\alpha_2$ form a vector-like pair so that when $n_1 = n_2$ (i.e. there are equal number of $\alpha_1$ and $\alpha_2$ fields) the 2d QFT is completely vector-like. In fact, for $n_1 = n_2$, one may identify $\Phi_1 = \Phi_2$ in the action and realize that the theory is just Witten's vector-like theory of superconducting string \cite{Witten:1984eb}, augmented by a coupling to  $U(1)_B$.  

For $n_1 \neq n_2$, there are non-vanishing anomalies from the bulk term. The 2d chiral anomalies from the string zero modes are given by \bea
&& \partial_\mu J_A^\mu  = - \frac{1}{4\pi} \left( n_1 - n_2 \right) \left( F_A^{(2)} + (q-n) F_B^{(2)} \right) ~,\\
&& \partial_\mu J_B^\mu  = - \frac{1}{4\pi} \left( n_1 - n_2 \right) \left( (q-n) F_A^{(2)} + (q-n)^2 F_B^{(2)} \right)~.
\eea 
Here, we can see that the 2d anomalies cancel the bulk anomalies exactly mod $n$ which, taking into account 
the gauge invariant local counterterms eq.~(\ref{eq:S_ct}),  leads to full anomaly cancellation from the bulk variation eq.~\eqref{eq:inflow_A} and \eqref{eq:inflow_B}.

\section{TQFT-Coupling II: Gauging a Discrete Subgroup}
\label{sec:TQFT-coupling II}

In this section, we discuss alternative ways to couple the axion-Maxwell theory to a TQFT.
In Section~\ref{subsec:discrete gauging_axion strings} 
 we consider a coupling of axion-Maxwell theory to a $\IZ_n$ BF theory which effectively gauges the $\IZ_n$ subgroup of axion 0-form shift symmetry. As we show below, this TQFT-coupling modifies the cosmic string spectrum in an interesting way, which can be in principle observable. This coupling also changes the 3-group symmetry structure as we show in Section~\ref{subsec:3-group_discrete_gauging}. We then further generalize this discussion to classify all possible TQFT-couplings via discrete gauging  in Section~\ref{subsec:other TQFT couplings}.

We begin with a charge $K$ axion-Maxwell theory 
\beq
S = \frac{1}{2} \int  d a \wedge * d a + \frac{1}{2 g^2} \int F^{(2)} \wedge * F^{(2)} - \frac{iK}{8\pi^2 f_a} \int a F^{(2)} \wedge F^{(2)}~.
\label{eq:Action_subgroup_gauging}
\eeq
The global symmetry structure of this simple theory is quite rich even without coupling to a TQFT. It is discussed in detail in \cite{Hidaka:2020iaz, Hidaka:2020izy, Brennan:2020ehu} and follows from the analysis of Appendix~\ref{app:symmetry_TQFT-coupling I} by removing all terms involving $K_B$ and $K_{AB}$. For reader's convenience we give a quick summary below. 

The symmetry structure of the charge $K$ axion-Maxwell theory is comprised of
\begin{itemize}
\item[1.] 0-form $\IZ_K^{(0)}$ axion shift symmetry: without the axion coupling to gauge fields, theory has a $U(1)^{(0)}$ shift symmetry $a \to a + c f_a, \; c \in \mathbb{S}^1$ with a corresponding current $*J_1 = i f_a * da$. The charge $K$ topological coupling 
breaks $U(1)^{(0)}\to\mathbb{Z}_K^{(0)}$.\footnote{As shown in \cite{Cordova:2022ieu,Choi:2022jqy}, the more correct statement is that the \emph{invertible} $U(1)$ shift symmetry gets converted into a set of \emph{non-invertible} shift symmetries. In particular, as far as correlation functions of local operators are concerned, the presence of non-invertible symmetry imposes selection rules that respect rational shift symmetries $\IQ/\IZ\subset U(1)^{(0)}$ (rather than $\mathbb{Z}_K\subset U(1)^{(0)}$). On the other hand, the non-invertible shift symmetry acts non-trivially on 't Hooft line operators. See \cite{Cordova:2022ieu,Choi:2022jqy} for more details and \cite{Cordova:2022fhg} for non-invertible symmetries in the Standard Model and beyond and novel applications in particle physics model building. \label{footnote:NIS}}
\item[2.]  2-form $U(1)^{(2)}$ axion winding symmetry: this symmetry is dual to the 0-form shift symmetry and 
 has the corresponding current $*J_3 = \frac{1}{2\pi f_a} d a$. This symmetry is a consequence of the Bianchi identity $d^2a=0$ and acts on 2d cosmic/axion strings.  

\item[3.] 1-form $\IZ_K^{(1)}$ electric symmetry: similar to the axion shift symmetry, pure Maxwell theory has a $U(1)^{(1)}$ electric 1-form symmetry following from the equation of motion $d*F^{(2)}=0$ which is then broken $U(1)^{(1)}\to \IZ_K^{(1)}$ by the charge $K$ topological coupling. 
This symmetry acts on Wilson line operators by shifting the gauge field by a discrete gauge field.\footnote{
Similar to the case of $U(1)^{(0)}$ symmetry discussed in Footnote \ref{footnote:NIS}, the invertible $U(1)^{(1)}$ electric symmetry is converted into a non-invertible $\IQ/\IZ$ symmetry. See \cite{Yokokura:2022alv,Choi:2022fgx} for more details. 
}
\item[4.] 1-form $U(1)^{(1)}$ magnetic symmetry: this symmetry is dual to the electric 1-form symmetry  and has the corresponding current $\ast J_2=\frac{1}{2\pi}F$. This symmetry is a direct consequence of the Bianchi identity $d F^{(2)} = 0$ and acts on `t Hooft lines which can be thought of as the IR-realization of massive, stable monopoles of the UV theory.
\end{itemize}
In the following, we will refer to $\IZ_K^{(0)},\IZ_K^{(1)}$ as ``electric symmetries'' -- since they shift local fields -- and the $U(1)^{(1)},U(1)^{(2)}$ as ``magnetic symmetries'' because they are both dual to an electric symmetry. 

In order to analyze these symmetries, we will couple to background gauge fields of the global symmetries listed above. The naive coupling leads to 
\bea
S = && \frac{1}{2} \int_{M_4} (da - f_a \mathcal{A}_e^{(1)} ) \wedge * (da - f_a \mathcal{A}_e^{(1)} ) + \frac{i}{2\pi f_a} \int_{M_4} da \wedge \mathcal{A}_m^{(3)} \nonumber \\
 + && \frac{1}{2 g^2} \int_{M_4} (F^{(2)} - \mathcal{B}_e^{(2)}) \wedge * (F^{(2)} - \mathcal{B}_e^{(2)}) + \frac{i}{2\pi} \int_{M_4} F^{(2)} \wedge \mathcal{B}_m^{(2)} \label{eq:BGF_subgroup_gauging_1} \\
 - && \frac{iK}{8\pi^2 f_a} \int_{N_5} (da - f_a \mathcal{A}_e^{(1)} ) \wedge (F^{(2)} - \mathcal{B}_e^{(2)}) \wedge (F^{(2)} - \mathcal{B}_e^{(2)})~, \nonumber
\eea
where, in the last line, we have written the axion-coupling on an auxiliary 5-manifold $N_5$ bounding our 4d spacetime $M_4$: $\partial
 N_5 = M_4$. This presentation makes the theory manifestly invariant under  background gauge transformations up to terms that are independent of the dynamical fields.\footnote{The additional terms that are generated by background gauge transformations that are independent of the dynamical fields can be interpreted as  `t Hooft anomalies. We will discuss these `t Hooft anomalies in Section \ref{subsec:3-group_discrete_gauging}.} 
  However, the above presentation is in fact dependent on the choice of $N_5$  due to the fact that background gauge fields $\mathcal{A}_e^{(1)}$ and $\mathcal{B}_e^{(2)}$ are $\mathbb{Z}_K$-valued, hence the integrand in the last line evaluated on a closed 5-manifold is not $2\pi \mathbb{Z}$. 
  
The theory can be made independent of $N_5$ by modifying the  ``magnetic'' symmetries so that instead of the standard $U(1)$-transformations, the background gauge fields $\CA_m^{(3)},\CB_m^{(2)}$ additionally transform under the electric symmetries: 
\bea
&& \mathcal{A}_m^{(3)} \to \mathcal{A}_m^{(3)} + d \lambda_m^{(2)} - \frac{K}{4\pi} \left( 2 \lambda_e^{(1)} \wedge \mathcal{B}_e^{(2)} + \lambda_e^{(1)} \wedge d \lambda_e^{(1)} \right)~, \label{eq:axion_MW_Am3_gauge_transf} \\
&& \mathcal{B}_m^{(2)} \to \mathcal{B}_m^{(2)} + d \lambda_m^{(1)} - \frac{K}{2\pi} \left( \lambda_e^{(0)} \mathcal{B}_e^{(2)} + \lambda_e^{(1)} \wedge \mathcal{A}_e^{(1)} + \lambda_e^{(0)} d \lambda_e^{(1)} \right)~.
\label{eq:axion_MW_Bm2_gauge_transf}
\eea

These modified symmetries then have modified field strengths $\mathcal{G}^{(4)},\mathcal{H}^{(3)}$:
\bea
&& \mathcal{G}^{(4)} = d \mathcal{A}_m^{(3)} + \frac{K}{4\pi} \mathcal{B}_e^{(2)} \wedge \mathcal{B}_e^{(2)}~, \label{eq:axion_MW_G4} \\
&& \mathcal{H}^{(3)} = d \mathcal{B}_m^{(2)} + \frac{K}{2\pi} \mathcal{A}_e^{(1)} \wedge \mathcal{B}_e^{(2)}~.
\label{eq:axion_MW_H3}
\eea
so that the action is written 
\bea
S = && \frac{1}{2} \int (da - f_a \mathcal{A}_e^{(1)} ) \wedge * (da - f_a \mathcal{A}_e^{(1)} ) + \frac{i}{2\pi f_a} \int a \mathcal{G}^{(4)} \nonumber \\
 + && \frac{1}{2 g^2} \int (F^{(2)} - \mathcal{B}_e^{(2)}) \wedge * (F^{(2)} - \mathcal{B}_e^{(2)}) + \frac{i}{2\pi} \int A^{(1)} \wedge \mathcal{H}^{(3)} \label{eq:BGF_subgroup_gauging_2} \\
 - && \frac{iK}{8\pi^2 f_a} \int_{N_5} (da - f_a \mathcal{A}_e^{(1)} ) \wedge (F^{(2)} - \mathcal{B}_e^{(2)}) \wedge (F^{(2)} - \mathcal{B}_e^{(2)})~, \nonumber
\eea
and is invariant under the choice of $N_5$. 
The modified transformation rules in eqs.~(\ref{eq:axion_MW_Am3_gauge_transf}) $\text{--}$ (\ref{eq:axion_MW_Bm2_gauge_transf}) show that the axion-Maxwell theory %
possesses a 3-group symmetry. 
Loosely speaking, an $n$-group is constructed from $0$-, $1$-,$\cdots$, $(n-1)$-form global symmetries where the $p$-form symmetries mix non-trivially with the $q$-form symmetries where $q>p$. In our case, eq.~(\ref{eq:axion_MW_G4}) shows that the $\mathbb{Z}_K^{(1)}$ symmetry turns on field strength of $U(1)^{(2)}$ symmetry and similarly, eq.~(\ref{eq:axion_MW_H3}) shows that the $\mathbb{Z}_K^{(0)}$, $\mathbb{Z}_K^{(1)}$ mix with the $U(1)^{(1)}$ symmetries. Note that there is a $\IZ_K^{(1)}$ that participates in the 3-group global symmetry and there is a $\IZ_{\sqrt{K}}^{(1)}\subset \IZ_K^{(1)}$ that decouples and is a genuine 1-form global symmetry. 
See Appendix~\ref{subapp:TQFT-coupling I_3-group} for more details.

\subsection{Gauging $\IZ_n^{(0)} \subset \mathbb{Z}_K^{(0)}$ and Axion-String Spectrum}
\label{subsec:discrete gauging_axion strings}

Since $\mathbb{Z}_K^{(0)}$ has no ABJ-anomaly, it is a good quantum symmetry and can be gauged\footnote{Here we assume that there is no cubic `t Hooft anomaly. This is allowed, but highly constrains the UV physics.}. Here, we will study the gauging of a subgroup $\IZ_n^{(0)} \subset \mathbb{Z}_K^{(0)}$ and its implications. 

First,  note that gauging the $\IZ_n$ discrete subgroup is equivalent to coupling the original theory to a $\IZ_n$ TQFT. Indeed, this is not specific to discrete gauging and is familiar from gauging continuous symmetries. Imagine a theory with a continuous global symmetry with a current $J_\mu$ coupled to a background gauge field $\mathcal{V}^\mu$
\beq
S \supset i \int J_\mu \mathcal{V}^\mu~.
\eeq
Gauging 
 this global symmetry is achieved by supplementing the theory with a dynamical gauge field. The above term then describes the coupling between the original theory and newly born gauge theory sector. In our case,  the gauge theory sector is a $\IZ_n$ TQFT and it couples to the theory analogously to the case of a continuous gauge field because 
 the $\IZ_n^{(0)}$ symmetry results from an explicitly broken $U(1)^{(0)}$ symmetry. 

In order to understand how the discrete gauging effects the axion theory, it is instructive to first  study the discrete gauging of free $U(1)$ Goldstone boson. 

\subsubsection{Discrete Gauging of Free $U(1)$ Goldstone Boson}
\label{subsubsec:discrete gauging_free GB}

Consider a $U(1)$-valued Goldstone boson $\phi$. This field has the standard action\footnote{We can alternatively write the action in terms of a canonically normalized field $\Phi=f\,\phi$ which satisfies $\Phi\sim \Phi+2\pi f$ analogously to the axion $a\sim a+2\pi f_a$. 
}
\eq{
S=\int \frac{f^2}{2}(d\phi)^2~, \qquad \phi\sim \phi+2\pi~.
}
This periodic identification of $\phi$ can be thought of as a gauge redundancy that transforms a naturally $\IR$-valued scalar field into a $U(1)$-valued scalar field. 
This theory has a $U(1)^{(0)}\times U(1)^{(2)}$ global symmetry corresponding to shift symmetry and winding symmetry. These symmetries have currents \eq{
\ast J_1=if^2\ast d\phi~, \qquad \ast J_3=\frac{d\phi}{2\pi} ~.
}
Here $\ast J_1$ is the momentum density -- generates shifts of $\phi$ -- and $\ast J_3$ measures the winding of $\phi$. The charged operators are $I (q, x)=e^{i q \phi(x)}$ and $\CS_2 (\ell, \Sigma_2)$ (which is the cosmic string operator of charge $\ell$) respectively. 

Now let us consider gauging the $\IZ_n^{(0)}\subset U(1)^{(0)}$ shift symmetry. In this case, we modify the action by coupling to the 1-form $\IZ_n$ gauge field $C^{(1)}$ 
\eq{
S=\int \frac{f^2}{2} (d\phi-C^{(1)}) \wedge * (d\phi-C^{(1)}) +\frac{i n}{2\pi}\int d C^{(1)}\wedge D^{(2)}~
\label{eq:discrete_gauging_free_GB}
}
where $D^{(2)}$ is a 2-form $\IZ_n$ gauge field. This gauging modifies the spectrum of operators in the theory in opposite ways: it projects out local operators (charged under $\IZ_n^{(0)}$) and adds ``dual'' string operators ($\IZ_n^{(2)}$ BF strings). 

First, note that the gauging identifies 
\eq{\label{phigauge}
\phi \sim \phi+\frac{2\pi}{n}~. 
}
This means that the gauging projects out local operators 
\eq{
I (q, x) , \quad q\notin n \IZ~,
}
because under the shift $\phi\mapsto \phi+\frac{2\pi}{n}$, the local operator shifts as
\eq{
I (q, x) \mapsto e^{\frac{2\pi i\, q}{n}} I (q, x)~, 
}
so that it is not gauge invariant unless $q\in n\IZ$. Alternatively, such a charged operator can be made a gauge invariant operator by attaching a (discrete) Wilson line to it 
\beq
I (q,x) \to \tilde{I} (q,x) = e^{iq\int_x^\infty C^{(1)} }I(q,x)~.
\eeq
For $q = n\mathbb{Z}$, however, the operator $\tilde{I} (q,x)$ becomes a well-defined \emph{local} operator because the attached Wilson line is trivial and $\tilde{I}(q,x)$ reduces to $I(q,x)$.\footnote{As discussed in detail in Appendix~\ref{app:review_BF}, the line operators $W (\Sigma_1) = e^{i \oint_{\Sigma_1} C^{(1)}}$ satisfy $W^n=1$ and so for $q\in n\mathbb{Z}$, one sees that $\tilde{I} (q,x)$ reduces to $I (q,x)$.}

Additionally, due to the identification \eqref{phigauge} we are now allowed to have \emph{fractional} winding numbers:
\eq{
\oint \frac{d\phi}{2\pi}\in \frac{1}{n}\IZ~.
}
These fractional winding numbers are in a sense ``added'' to the spectrum -- they are constructed by $\phi$ passing through $2\pi/n$ of the $2\pi$ period which is then accompanied by a $\IZ_n$ gauge transformation. This winding $\IZ_n$ gauge transformation requires a non-trivial gauge background given by an insertion of  the BF string operator
\eq{
W_2 (\ell, \Sigma_2)=e^{i \ell \oint_{\Sigma_2} D^{(2)}}
}
which are charged under the 2-form $\IZ_n^{(2)}$ global symmetry that shifts by $D^{(2)} \to D^{(2)} + \frac{2\pi}{n} \lambda_2$ with $\oint \lambda_2 \in 2\pi \mathbb{Z}$.  These string operators are classified by the $\IZ_n$ charges $\ell=0,1,\cdots, (n-1)$.

We can see that the insertion of the BF string operator leads to the fractional $\phi$-winding configurations explicitly by noting that in the presence of the BF string the equations of motion for $C^{(1)}$ are given by:
\eq{
d C^{(1)} = \frac{2\pi\, \ell}{n}\delta^{(2)}(\Sigma_2)
}
which implies that $\oint \frac{C^{(1)}}{2\pi}=\frac{\ell}{n}$ in the presence of the BF string. Now we can use the fact that the combination $d\phi-C^{(1)}$ is $\IZ_n$-gauge invariant to see that in the presence of the BF string 
\eq{
\oint \frac{C^{(1)}}{2\pi}=\oint \frac{d\phi}{2\pi}~{\rm mod}_\IZ=\frac{\ell}{n}~. 
}
See Appendix~\ref{app:review_BF} for a detailed discussion. 

As far as local physics are concerned, this theory is IR equivalent to the original free periodic scalar. 
This may be seen explicitly as follows. Consider rescaling: $\phi \to \tilde{\phi} = n \phi, \; C^{(1)} \to \tilde{C}^{(1)} = n C^{(1)}, \; f \to \tilde{f} = f/ n$. It is easy to show that the theory with discrete gauging eq.~(\ref{eq:discrete_gauging_free_GB}) becomes
\eq{
S=\int \frac{\tilde{f}^2}{2} (d \tilde{\phi}- \tilde{C}^{(1)}) \wedge * (d \tilde{\phi}-\tilde{C}^{(1)}) +\frac{i}{2\pi}\int d \tilde{C}^{(1)}\wedge D^{(2)}~.
\label{eq:discrete_gauging_free_GB_rescaled}
}
Now the second term trivializes the $\tilde{C}^{(1)}$ and $D^{(2)}$ gauge fields (it is a $\IZ_n$ gauge theory with $n=1$) so that effectively none of the $U(1)$ shift symmetry of $\tilde{\phi}$ is gauged. Accordingly, $\tilde{C}^{(1)}$ can be  removed from the theory and then one can easily check that $\tilde{\phi}$ has $2\pi$ periodicity and its winding number is given by $\oint d\tilde{\phi} \in  2\pi \mathbb{Z}$. 

One may then wonder if $\IZ_n \subset U(1)$ discrete gauging has any physical consequences. The answer is: discrete gauging does have important physical implications and in current example it is captured by cosmic string physics. We have shown that in addition to the original global strings, the discrete gauging adds $\IZ_n$-classified local (BF) strings. While the tension of the former is of the order $\sim f^2$, the tension of the latter is the square of the scale at which the BF theory emerges $\sim\Lambda_{\rm BF}^2$. In general, this second scale can be parametrically larger than $f$ and therefore, the varying tension of the cosmic string spectrum can encode the effect of $\IZ_n$ gauging.

\subsubsection{Discrete Gauging of Axion-Maxwell Theory}
\label{subsubsec:discrete gauging Axion-Maxwell}

Now let us return to axion-Maxwell theory and discuss the partial gauging of discrete $\IZ_n^{(0)} \subset \IZ_K^{(0)}$. After gauging, the action becomes 
\bea
S & = & \frac{1}{2} \int  (d a - f_a C^{(1)}) \wedge * ( d a - f_a C^{(1)} ) + \frac{1}{2 g^2} \int F^{(2)} \wedge * F^{(2)} \nonumber \\ 
&& + \frac{iK}{f_a} \int (d a - f_a C^{(1)}) \wedge \omega_3 (A^{(1)}) + \frac{i n}{2\pi} \int D^{(2)} \wedge d C^{(1)}
\label{eq:action_Z_n_gauged}
\eea
where $C^{(1)}$ is the dynamical gauge field for the $\IZ_n^{(0)}$ and $\omega_3 (A^{(1)})$ is the 3d $U(1)$ Chern-Simons action as defined in Section~\ref{subsec:anomaly inflow}. Also, $D^{(2)}$ is the 2-form gauge field within the $\IZ_n$ BF theory sector.

It is clear from the definition of the axion shift symmetry, that $\IZ_K^{(0)}$ is a non-linearly realized discrete symmetry and the gauging, therefore, corresponds to gauging a Higgsed discrete symmetry. This is in general a result of the fact that the axion is the pseudo-Goldstone field for a spontaneously broken (anomalous) $U(1)$ symmetry of which $\IZ_K^{(0)}\subset U(1)^{(0)}$ is a non-anomalous subgroup. 

However, even though we are gauging the axion shift symmetry, we still have a continuous field with dynamical excitations. While gauging of $U(1)$ global symmetry would have removed all of the fluctuations of periodic scalar, here the gauging is applied only to a discrete subgroup and it turns only a ``measure-zero'' part into gauge degrees of freedom, leaving most of the continuous excitations intact.

As in the case of the free periodic scalar, the discrete gauging removes some of the local operators and increases the number of surface operators. First, consider the allowed local operators. Without gauging, the charged objects under the 0-form $\mathbb{Z}_K^{(0)}$ is local operators $I(q,x) = e^{i q \,a(x)/f_a}$ with a charge $q\in \mathbb{Z}$ (recall that the axion is $2\pi f_a$-periodic). It shifts as
\beq
I (q,x) = e^{i q\, a(x)/f_a } \to e^{\frac{2\pi i}{K} q} I(q,x)
\eeq
under global $\mathbb{Z}_K^{(0)}$ transformations. As in the case of the free periodic scalar, after gauging $\IZ_n^{(0)} \subset \mathbb{Z}_K^{(0)}$, these local operators are no longer gauge invariant for general $q \in \mathbb{Z}$. Rather, they are only gauge invariant only for $q \in n \mathbb{Z}$ -- otherwise they require attaching a Wilson line and become non-local.

Now let us consider the non-local string-like operators. 
Before gauging $\IZ_n^{(0)}$, the axion is a periodic scalar with the period $a \sim a + 2\pi f_a$.  After gauging $\IZ_n^{(0)}$, the $\IZ_n$ gauge transformations shift $a \sim a + \frac{2\pi f_a}{n}$. This allows axion field configurations with  fractional  winding number
\beq
\oint \frac{d a}{f_a} \in \frac{2\pi}{n} \mathbb{Z}
\eeq
 because they are now gauge equivalent to winding configurations that traverse a complete period. Cosmic strings with integral winding are global axion strings, and the ones with fractional $\IZ_n$-valued windings are BF strings. As discussed in the previous section, this can be seen explicitly by noting that the equation of motion for $a$ implies that
\beq
\oint \frac{da}{2\pi f_a} {\rm mod}_{\mathbb{Z}} = \oint \frac{C^{(1)}}{2\pi} \in \frac{1}{n} \mathbb{Z},
\eeq
which is activated by an insertion of BF surface operator $W_2 (\Sigma_2, m) = e^{i m \oint_{\Sigma_2} D^{(2)}}$. 

 At this point we recall that the scale at which the $\IZ_n$ gauge theory emerges,\footnote{Here we insist on the existence of a UV completion which only contains continuous gauge symmetries. The scale $\Lambda_{\rm BF}$ is the scale at which the continuous gauge symmetry is broken to $\IZ_n$. } $\Lambda_{\rm BF}$, should be generically higher than the scale where the axion is emerges from a spontaneously broken, anomalous $U(1)$ symmetry, $f_a$.  The reason for the hierarchy $\Lambda_{\rm BF}\gtrsim f_a$ is simply because, in any UV completion, 
a discrete global symmetry in the low energy EFT (non-linearly realized or otherwise) can only be gauged if the 
gauge field is also discrete at or above the scale where the symmetry emerges:  $\Lambda_{\rm sym}\gtrsim \Lambda_{\rm EFT}$.  Therefore, one expects on a general grounds that $\Lambda_{\rm BF} \gtrsim f_a$. 
 
 This hierarchy has a measurable effect on the spectrum of cosmic strings in the theory. In the $\IZ_n^{(0)}$-gauged theory, cosmic strings consists of global axion strings with tension $T\sim f_a^2$ and $\mathbb{Z}$-valued winding numbers, and BF strings with tension $T\sim \Lambda_{\rm BF}^2$ and ``winding numbers'' $m \in \frac{1}{n} \times ( 1, 2, \cdots, (n-1) )$.\footnote{ The local strings of BF theory are not really defined by a winding number as they do not possess any topological charge \cite{Polchinski:2005bg}. Rather, they are measured by a conserved magnetic flux $m=\frac{1}{2\pi}\oint C^{(1)}$. In an Abelian Higgs model, however, we have $\oint (d\varphi -  C^{(1)}) \in \IZ$ and the magnetic flux is transfered to the would-be Goldstone boson as a winding number. }

\subsection{UV Field Theory Completion}
\label{subsec:UV theory_TQFT coupling II}

In this section, we show that the axion-Maxwell theory coupled to a TQFT as in the previous section can arise as a long distance description of a local QFT. This result  demonstrates that TQFT couplings and their associated remarkable features can appear rather ubiquitously in a broad class of particle physics models. 

In order to formulate a UV completion, we recall that coupling to the TQFT can be thought of as gauging a discrete subgroup of the axion shift symmetry. If we take the KSVZ UV completion of axion-Maxwell theory, the discrete gauging corresponds to gauging $\IZ_n^{(0)} \subset U(1)_{\rm PQ}$. We can then try to couple to the $\IZ_n$ TQFT via an Abelian Higgs Model which breaks a gauge symmetry $U(1)_B \to \IZ_n$. Requiring that the emergent $\IZ_n$ gauge symmetry act on $U(1)_{\rm PQ}$ requires that we have the UV symmetry structure:
\beq
G_{\rm UV}=
U(1)_A \times \frac{U(1)_B \times U(1)_{\rm PQ}}{\IZ_n}
\label{eq:TQFT coupling II_symmetry overlap}
\eeq
where we used the same notations for gauge groups as in Section~\ref{sec:TQFT-coupling I}: $U(1)_A$ for unbroken electromagnetism and $U(1)_B$ for the gauge group that is spontaneously broken $U(1)\to \IZ_n$. The second factor means that $\IZ_n \subset U(1)_{\rm PQ}$ is redundant and can be undone by means of $\IZ_n \subset U(1)_B$ rotations. This means that the true global symmetry of the theory is given by projecting out the $\IZ_n$ gauge symmetry from the $U(1)_{\rm PQ}$ so that 
\eq{G_{\rm global}=U(1)_{\rm PQ}/\IZ_n~.}
Below, we will present a theory that realizes this symmetry structure. 
\begin{table}
\center
\begin{tabular}{|c||c|c|c|c|}
\hline
  & $U(1)_A$ & $U(1)_B$ &  $U(1)_{\rm PQ}$ & $U(1)_F$ \\
\hline\hline
$\psi_+$ & $1$ & $0$ & $-n$ & $n$\\
$\psi_-$ & $-1$ & $-1$ & $-1$ & $-1$\\
\hline
$\chi_+$ & $1$ & $n+1$ & $n+1$ & $n+1$ \\
$\chi_-$ & $-1$ & $0$ & $-n$ & $-n$\\
\hline
$\eta_+$ & $1$ & $1$ & $1$ & $1$\\
$\eta_-$ & $-1$ & $-n-1$ & $-n-1$ & $-n-1$\\
\hline
$\Phi_1$ & $0$ & $1$ & $n+1$ & $-n+1$\\
$\Phi_2$ & $0$ & $-n-1$ & $-1$ & $-1$\\
$\Phi_3$ & $0$ & $n$ & $n$ & $n$\\
\hline
\end{tabular}
\caption{Quantum numbers of the fields of the UV completion eq.~(\ref{eq:UV_theory_2}). Below $E < \langle \Phi_I \rangle$, this theory matches to our effective theory. 
} 
\label{tab:charges2}
\end{table}

Consider a theory whose Lagrangian is 
\bea
\mathcal{L}_{\rm UV}= && - \frac{1}{4 g_A^2} F_A^{(2)} \wedge * F_A^{(2)} - \frac{1}{4 g_B^2} F_B^{(2)} \wedge * F_B^{(2)} +  \sum_{k=\pm} \bar{\psi}_k i \slashed{D} \psi_k + \bar{\chi}_k i \slashed{D} \chi_k + \bar{\eta}_k i \slashed{D} \eta_k \nonumber  \\
&& + |D \Phi_1|^2 + |D \Phi_2|^2 +|D \Phi_3|^2 - V(\Phi_1, \Phi_2,\Phi_3)\nonumber \\
&& - \lambda_{1} \Phi_1 \psi_+ \psi_- - \lambda_{2} \Phi_2 \chi_+ \chi_- - \lambda_{3} \Phi_2 \eta_+ \eta_- + {\rm h.c.} ~.
\label{eq:UV_theory_2}
\eea
The scalar potential is such that both $\langle \Phi_1 \rangle$, $\langle \Phi_2\rangle$, and $\langle \Phi_3 \rangle$ are non-zero. 

We will assign gauge and global symmetry charges to the fields which are summarized in Table~\ref{tab:charges2}.
These choices of quantum number are tightly constrained by multiple requirements:

\begin{itemize}
\item[(i)] In order to construct a well defined UV theory, we first choose the charges of the gauge symmetries $U(1)_A$ and $U(1)_B$ so that they are anomaly free. 

\item[(ii)] In order to produce an axion, we demand that $U(1)_{\rm PQ}$ is the only spontaneously broken global symmetry with any ABJ anomalies and that it only has an ABJ anomaly with $U(1)_A$. 
The ABJ anomalies in our model for $U(1)_{\rm PQ}$ are given by\footnote{We can additionally obtain a theory with axion-Maxwell coupling $K=-2n p$ by taking $p$-copies of the fermion sector without increasing the number of scalar fields. This increases the number of vector-like symmetries that are not spontaneously broken in the IR, while increasing the size of the $U(1)_{\rm PQ}\times [U(1)_A]^2$ ABJ anomaly. }
\eq{
U(1)_{\rm PQ}\left[U(1)_{A}\right]^2
=-2n
~, \quad
U(1)_{\rm PQ}\left[U(1)_{B}\right]^2=
U(1)_{\rm PQ}\,U(1)_{A}\,U(1)_B
=0~.
}
By a simple counting argument, there are three remaining $U(1)$ global symmetries which are parametrized by $U(1)_F$ and two vector-like symmetries $U(1)_{V_1}$ and $U(1)_{V_2}$ (not present in Table \ref{tab:charges2}). The vector-like symmetries $U(1)_{V_i}$ act only on the fermions and hence are not spontaneously broken by Higgs condensation. The global symmetry $U(1)_F$ is spontaneously broken but has vanishing ABJ anomalies. 

\item[(iii)] In order to realize non-trivial overlap described by $\frac{U(1)_B \times U(1)_{\rm PQ}}{\IZ_n}$, we need the charges of $U(1)_{\rm PQ}$ to reduce to those of $U(1)_B~{\rm mod}_n$. 
This can be easily verified upon inspection of Table \ref{tab:charges2}. 
\end{itemize}

\noindent The theory described by the Lagrangian in eq.~(\ref{eq:UV_theory_2}) with charge assignments in Table~\ref{tab:charges2} manifestly has the symmetry structure in eq. \eqref{eq:TQFT coupling II_symmetry overlap}.

To reproduce the axion-Maxwell coupled to a $\IZ_n$ TQFT in the IR, we choose the parameters of the scalar potential such that $\langle \Phi_3\rangle\gg\langle \Phi_{1,2}\rangle$. At intermediate energies $E$, where $\langle\Phi_3\rangle\gg E\gg\langle \Phi_{1,2}\rangle$, $U(1)_B$ is broken by $\langle \Phi_3\rangle$ to $U(1)_B\to \IZ_n$  and the theory flows to a KSVZ model coupled to a $\IZ_n$ TQFT where $U(1)_{\rm PQ}$ is gauged by $\IZ_n\subset U(1)_B$. This intermediate theory has string-defects corresponding to the $\IZ_n$ BF strings which come from the unscreened $\Phi_3$ Nielsen-Olesen vortices. 
Then, in the deep IR $E\ll\langle \Phi_{1,2}\rangle$, the theory flows to charge $2n$ axion-Maxwell theory coupled to a $\IZ_n$ gauge field and a ``decoupled'' $U(1)$ Goldstone boson where the axion and auxiliary Goldstone boson has charge 1 under $\IZ_n $. Note that we can add higher dimension operators that break $U(1)_F$ explicitly so that we land exactly on our IR theory. 

In this UV completion, we find that $\Lambda_{\rm BF}=\langle \Phi_3\rangle$ and $f_a\sim\langle \Phi_{1,2}\rangle$ and that we respect the hierarchy $\Lambda_{\rm BF}\gtrsim f_a$. However, there are other limits of this UV theory we can consider -- for example by trying to ``invert'' the hierarchy. We will find that when we invert the hierarchy, the RG flow will lead to a different IR theory, thus side-stepping the constraints from EFT considerations. 

Let us first consider the limit $\langle \Phi_1\rangle\gg \langle \Phi_{2,3}\rangle$. At intermediate energies $E$, where $\langle \Phi_1\rangle\gg E\gg \langle \Phi_{2,3}\rangle$, $U(1)_B$ is completely broken. Now the intermediate theory is given by a KSVZ theory. In this KSVZ theory, the $U(1)_{\rm PQ}$ and $U(1)_F$ symmetries have become degenerate as shown in Table \ref{tab:Phi1condense}

\begin{table}
\begin{center}
\begin{tabular}{|c||c|c|cc|}
\hline
  & $U(1)_A$  &  $U(1)_{{\rm PQ},F}$ & $U(1)_{\widetilde{PQ}}$ &$U(1)_{\widetilde{F}}$\\
\hline\hline
$\chi_+$ & $1$  & $n+1$ & $1$ & $1$\\
$\chi_-$ & $-1$  & $-n$ & $0$ & $0$\\
\hline
$\eta_+$ & $1$ & $1$ & $0$ & $0$\\
$\eta_-$ & $-1$  & $-n-1$ & $1$ & $-1$\\
\hline
$\Phi_2$ & $0$  & $-1$ &  $-1$ & $-1$\\
$\Phi_3$ & $0$  & $n$ &$-1$ & $1$\\
\hline
\end{tabular}\end{center}
\caption{The quantum numbers of the fields in the theory at intermediate energy levels $\langle \Phi_1\rangle\gg E \gg \langle \Phi_{2,3} \rangle$. Here $U(1)_{\rm PQ},U(1)_{F}$ have become degenerate. 
} 
\label{tab:Phi1condense}
\end{table}
Instead, since the phase of $\Phi_1$ is completely eaten by $U(1)_B$, we must  redefine the global symmetries  $U(1)_{\rm PQ},U(1)_F$ so that they are orthogonal to the broken gauge symmetry, such as $U(1)_{\widetilde{PQ}},U(1)_{\widetilde{F}}$ 
with charges defined in Table \ref{tab:Phi1condense}.\footnote{
Note that this differs from the case where $\langle \Phi_3\rangle\gg \langle \Phi_{1,2}\rangle$ where the $\IZ_n\subset U(1)_B$ gauge symmetry, which overlaps with $\IZ_n\subset U(1)_{\rm PQ}\times U(1)_F$. In this case, we are not required to project onto the orthogonal symmetry because of the overlapping, preserved gauge symmetry. 
} 
The reason we should redefine the global symmetries is that by breaking  $U(1)_B$, we are required to implicitly project onto orthogonal global symmetries so that the transformations of the remaining fields do not move along the broken, gauged direction.

Now consider the other limit where $\langle \Phi_2 \rangle \gg \langle \Phi_{1,3} \rangle$. 
At intermediate energies $E$, where $\langle \Phi_2 \rangle\gg E\gg \langle \Phi_{1,3}\rangle$, $U(1)_B$ is broken down to $\IZ_{n+1}$ and the intermediate theory is described by a KSVZ model coupled to a $\IZ_{n+1}$ BF TQFT. The charges of the resulting KSVZ model are given in Table \ref{tab:Phi2condense}. 

Due to the symmetry structure: 
\eq{
G_{\rm total}=U(1)_A\times \frac{U(1)_B\times U(1)_{\rm PQ}\times U(1)_F}{\IZ_n\times\IZ_n}
}
we know that $\langle \Phi_2\rangle$ breaks the symmetry to 
\eq{
G_{\rm total}\to U(1)_A\times \IZ_{n+1}\times \frac{U(1)_{\rm PQ}\times U(1)_F}{\IZ_n}
}
and the $\IZ_{n+1}$ gauge symmetry can be decoupled from $U(1)_{\rm PQ},U(1)_F$. Again, we see that breaking the continuous part of $U(1)_B$ demands that we 
project $U(1)_B$ out of the global symmetries. This can be accomplished by 
the modified $U(1)_{\widetilde{PQ}},U(1)_{\widetilde{F}}$ symmetries generated by 
\eq{
Q_{\widetilde{PQ}}=Q_{\rm PQ}-Q_A~, \qquad Q_{\widetilde{F}}=Q_{F}-Q_A 
}
which completely decouples from $\IZ_{n+1}$. Thus, when we flow to the deep IR $\IZ_{n+1}$ is spontaneously broken by $\langle \Phi_{1,3}\rangle$, and there is a remaining axion with charge 2 and decoupled $U(1)$ Goldstone boson. Therefore in our UV completion, we find that if we tune the scales of the theory so that  $\sigma=\Lambda_{\rm BF}/f_a\to 0$, that there will be a phase transition near $\sigma\approx 1$ where the IR theory will have different IR dynamics as described above.

\begin{table}
\begin{center}
\begin{tabular}{|c||c|c|c|c|}
\hline
  & $U(1)_A$ & $\IZ_{n+1}$ &  $U(1)_{\rm PQ}$ & $U(1)_F$ \\
\hline\hline
$\psi_+$ & $1$ & $0$ & $-n$ & $n$\\
$\psi_-$ & $-1$ & $-1$ & $-1$ & $-1$\\
\hline
$\eta_+$ & $1$ & $1$ & $1$ & $1$\\
$\eta_-$ & $-1$ & $0$ & $-n-1$ & $-n-1$\\
\hline
$\Phi_1$ & $0$ & $1$ & $n+1$ & $-n+1$\\
$\Phi_3$ & $0$ & $n$ & $n$ & $n$\\
\hline
\end{tabular}\end{center}
\caption{The quantum numbers of the fields in the theory at intermediate energy levels $\langle \Phi_2\rangle\gg E \gg \langle \Phi_{1,3} \rangle$. } 
\label{tab:Phi2condense}
\end{table}

Note that we can physically think of the $\IZ_n$ gauging as a being a consequence of the fact that we ``chose our symmetries'' in a particular way. To see this, note that if we lift condition (iii) above, it is more natural to parametrize the global symmetries by 

\begin{center}\begin{tabular}{c|cccccc|ccc}
& $ \psi_+ $ &  $ \psi_- $ & $ \chi_+ $ & $ \chi_- $ & $ \eta_+ $ & $ \eta_- $ & $ \Phi_1 $ & $ \Phi_2 $ & $ \Phi_3 $ \\
\hline
$U(1)_1 $ &1& $ -1 $ &0&0&0&0&0&0&0\\
 $ U(1)_2 $ &0&0&1& $ -1 $ &0&0&0&0&0\\
 $ U(1)_3 $ &1&0&0&1&0&0& $ -1 $ & $ -1 $ &0\\
 $ U(1)_4 $ &1&0&0& $ -1 $ &0&0& $ -1 $ &1&0
\end{tabular}\end{center}
Here $U(1)_{1,2}$ correspond to $U(1)_{V_1,V_2}$ which are not spontaneously broken when the $\Phi_I$ condense. The global symmetries $U(1)_{\rm PQ}$ and $U(1)_F$ (generated by $Q_{\rm PQ}$ and $Q_F$ respectively) are related to $U(1)_{B}$ and $U(1)_{3,4}$ (generated by $Q_B,Q_{3,4}$) as 
\eq{
Q_F=Q_B+n Q_3\quad, \quad Q_{\rm PQ}=Q_B+n Q_4~.
}
In other words, we have chosen $U(1)_{\rm PQ}$ and $U(1)_F$ so that they overlap with $U(1)_B$ mod$_n$ by construction -- this redundancy is in some sense artificially engineered. 

In this construction, it is clear how the $\IZ_n\subset U(1)_B$ gauge symmetry effects the resulting IR theory. First, note that after condensing $\Phi_3$,  the charges of remaining fields under $\IZ_n$ are given by 

\begin{center}\begin{tabular}{c|cccc|cc}
& $ \psi_+ $ &  $ \psi_- $ & $ \chi_+ $ & $ \chi_- $ & $ \Phi_1 $ & $ \Phi_2$\\
\hline
$\IZ_n $ &0& $ -1 $ &1&0&1& $-1$
\end{tabular}\end{center}
By shifting by $\IZ_n\subset U(1)_A$, we can see that $\IZ_n$ gauges $U(1)_4$ and does not act on the $U(1)_3$.  This implies that the $\IZ_n$ gauging can be decoupled from the axion by a field redefinition although the BF strings are indeed physical. 

In summary, we find that in the parameter space spanned by the vevs $\langle \Phi_{1,2,3}\rangle$, there is only one hierarchy that flows to the axion-Maxwell theory coupled to a $\IZ_n$ BF theory. 
This hierarchy of scales is given by $\langle \Phi_3\rangle\gg \langle \Phi_{1,2}\rangle$ and leads to $\Lambda_{\rm BF}\gg f_a$, thus reproducing the expectation from EFT and higher-group symmetry considerations.

\subsection{3-Group}
\label{subsec:3-group_discrete_gauging}

We now return to the 3-group symmetry structure of axion-Maxwell coupled to the $\IZ_n$ TQFT and its associated `t Hooft anomalies. Understanding the higher-group symmetry structure of this theory is essential to understanding all gaugable symmetries, hence all possible non-trivial TQFT-couplings that arise from discrete gauging. We will 
discuss the set of possible gaugings in the next section. 
Here, we will focus on the 3-group symmetry of charge $K$ axion-Maxwell theory and the effect of $\IZ_n^{(0)} \subset \mathbb{Z}_K^{(0)}$ gauging. 

To this end, let us first review `t Hooft anomalies in theory without discrete gauging. 't Hooft anomalies can be probed by turning on background gauge fields for the global symmetries and studying the behavior of the partition function under their associated gauge transformations. In particular, we say that a symmetry has an `t Hooft anomaly if the path integral is not invariant under the associated background gauge transformations when all background gauge fields are turned on:
\eq{
Z[A_I+d\lambda_I]=e^{-i \int \CA^{(1)}[\{A_I\};\lambda_I]}\,Z[\{A_I\}]\quad \Longleftrightarrow \quad \{\text{`t Hooft anomaly}\}
}
where here $\CA^{(1)}[\{A_I\};\lambda]$ captures the anomalous variation of the action which is only dependent on the background gauge fields $\{A_I\}$. 
In general, the anomalous variation $\CA^{(1)}$ can be described by inflow of the variation of a 5d ``anomaly action'' $\CA[\{A_I\}]$ following the ``descent procedure''. 

In axion-Maxwell theory, the theory contains couplings of the form 
\eq{
 S =...+ \frac{i}{2\pi f_a} \int_{M_4} a\, \mathcal{G}^{(4)} + \frac{i}{2\pi} \int_{M_4}A^{(1)} \wedge \mathcal{H}^{(3)} ~.%
}
These terms are not invariant under electric background gauge transformations and generate transformations of the form 
\eq{
\delta S=\frac{i}{2\pi }\int_{M_4}\lambda_e^{(0)} \CG^{(4)}+\frac{i}{2\pi}\int_{M_4} \lambda^{(1)}_e\wedge \CH^{(3)}~.
} 
Since the background gauge fields transform as $\delta \mathcal{A}_e^{(1)} = d \lambda_e^{(0)}$ and $\delta \mathcal{B}_e^{(2)} =d \lambda_e^{(1)}$, the anomalous variation is captured by the 5d anomaly inflow action
\bea
&& S_{\rm inflow} = \frac{i}{2\pi} \int_{N_5} \mathcal{A}_e^{(1)} \wedge \mathcal{G}^{(4)} + \mathcal{B}_e^{(2)} \wedge \mathcal{H}^{(3)} \label{eq:subgroup_gauging_inflow_action} \\
&& \hspace{1.13cm} = \frac{i}{2\pi} \int_{N_5} \mathcal{A}_e^{(1)} \wedge \left( d \mathcal{A}_m^{(3)} + \frac{K}{4\pi} \mathcal{B}_e^{(2)} \wedge \mathcal{B}_e^{(2)} \right) + \mathcal{B}_e^{(2)} \wedge \left( d \mathcal{B}_m^{(2)} + \frac{K}{2\pi} \mathcal{A}_e^{(1)} \wedge \mathcal{B}_e^{(2)} \right). \nonumber
\label{Anomalyinflowtype2}
\eea
The first term describes a mixed anomaly between $\mathbb{Z}_K^{(0)}$ and $U(1)^{(2)}$, which due to the 3-group symmetry, also leads to a mixed anomaly between $\mathbb{Z}_K^{(0)}$ and $\mathbb{Z}_K^{(1)}$. Similarly, the second term describes a mixed anomaly between $\mathbb{Z}_K^{(1)}$ and $U(1)^{(1)}$, which also leads to a mixed anomaly between $\mathbb{Z}_K^{(0)}$ and $\mathbb{Z}_K^{(1)}$  due to the 3-group symmetry.

Now let us consider the effect of gauging $\IZ_n^{(0)} \subset \mathbb{Z}_K^{(0)}$.  Here, the background gauge field $\mathcal{A}_e^{(1)}$ decomposes into a dynamical gauge field $C^{(1)}$ for the gauged $\IZ_n^{(0)}$ part a background gauge field (which by abuse of notation we also call $\CA_e^{(1)}$) for the $\IZ_{K/n}^{(0)}$ part. Now, the mixed anomalies involving the dynamical gauge field $C^{(1)}$ can be viewed as ABJ-type anomalies. 

First, let us consider the effect of the gauging on $U(1)^{(2)}$. To study this, let us consider turning off $\CB_e^{(2)},\CB_m^{(2)}$. Now we can write the anomaly action as 
\eq{
S_{\rm inflow}=-\frac{i}{2\pi}\int_{N_5}d\CA_e^{(1)}\wedge \CA_m^{(3)}~.
}
In this case, we find that the anomalous variation of the action is given by 
\eq{
\delta S_{\rm inflow}=-\frac{i}{2\pi}\int_{M_4}\CA_e^{(1)}\wedge d\Lambda_m^{(2)}~, \qquad \delta \CA_m^{(3)}=d\Lambda_m^{(2)}~. 
}
This implies that gauging $\IZ_n^{(0)}$ extends the  periodicity of $U(1)^{(2)}$  to $2\pi n$. This is a consequence of the fact that $U(1)^{(2)}$ is the dual symmetry of $\IZ_{K/n}^{(0)}\subset U(1)/\IZ_n^{(0)}\cong U(1)^{(0)}$. This is analogous to the fact that rescaling the periodicity of a periodic scalar (i.e.~gauging) as we saw before  also rescales the quantum numbers of the momentum and winding states oppositely. 

Now let us consider the consequence of the $\IZ_n$ gauging on the $\IZ_K^{(1)}$ global symmetry. The correct way to show that the $\IZ_K^{(1)}$ symmetry is modified is to first choose counterterms so that the action is invariant under $\IZ_n$-gauge transformations. Then, we see that the variation of the action from inflow is given by 
\eq{
\delta S_{\rm inflow}=\frac{iK}{4\pi}\int \CA_e^{(1)}\wedge \left(2\lambda_e^{(1)}\wedge \CB_e^{(2)}+\lambda_e^{(1)}\wedge d\lambda_e^{(1)}\right)~. 
}
Because the $\IZ_n$ gauging sums over $\CA_e^{(1)}$ with $\oint \CA_e^{(1)}\in \frac{2\pi}{n}\IZ$,  we see that the $\IZ_K^{(1)}$ is broken  $\IZ_K^{(1)}\to \IZ_{K/n}^{(1)}$.\footnote{Using the fact that $\oint\frac{\CB^{(2)}}{2\pi}\in \frac{1}{K}\IZ$ for $\CB^{(2)}$ a $\IZ_K$ 2-form gauge field, we see that charge $K$ axion theory has a genuine $\IZ_p^{(1)}$ 1-form global symmetry for $K=mp^2$. In the case where we gauge $\IZ_n^{(0)}\subset \IZ_K^{(0)}$, the $\IZ_p^{(1)}$ is broken $\IZ_p^{(1)}\to \IZ_q^{(1)}$ for $K=k nq^2$. This can be seen in our UV model from the fact that we can introduce the 1-form electric symmetry in the UV iff we assign the $U(1)_A$ charges $\pm q$ to the fermion pairs which produces $K=2nq^2$. } Note that the gauging of $\IZ_K^{(0)}$ does not affect the $U(1)^{(1)}$ magnetic symmetry since $\IZ_{K/n}^{(1)}$ has trivial pairing with $\IZ_n^{(0)}$ in the $\CB_m^{(2)}$ transformation laws. 

We thus find that the symmetry of the theory after $\IZ_n^{(0)}$-gauging is a 3-group consisting of $\IZ_{K/n}^{(0)}$, $U(1)^{(1)}_m$, $\IZ_{K/n}^{(1)},$ $U(1)^{(2)}$ which describes the global symmetry group of charge $K/n$ axion-Maxwell theory.

\subsection{Other TQFT Couplings via Discrete Gauging}
\label{subsec:other TQFT couplings}

As we have seen, one way to couple a theory to a TQFT is by a gauging a discrete symmetry. It is clear from our analysis above that this process is non-trivial and can lead to interesting features \cite{Kapustin:2014gua}. Given a theory described by a local QFT, potential TQFT couplings of this sort can be systematically analyzed by studying the theory's higher-group global symmetry and associated 't Hooft anomalies. Here, we demonstrate this analysis on charge $K$ axion-Maxwell theory as a concrete example. 

As we have discussed in the previous section, charge $K$ axion-Maxwell theory posses a 3-group global symmetry structure. This 3-group involves $U(1)^{(2)}$ axion winding symmetry that is intertwined with $\mathbb{Z}_K^{(1)}$ electric symmetry and a $U(1)^{(1)}$ magnetic symmetry that is interlaced with $\mathbb{Z}_K^{(1)}$, and $\mathbb{Z}_K^{(0)}$ axion shift symmetry. The 't Hooft anomalies of this theory are described the 5d inflow action eq.~(\ref{eq:subgroup_gauging_inflow_action}) which describes a mixed anomaly between $\mathbb{Z}_K^{(0)}$ and the mixed $U(1)^{(2)}-\IZ_K^{(1)}$ symmetry as well as a mixed anomaly between $\mathbb{Z}_K^{(1)}$ and the interlaced $U(1)^{(1)}-\IZ_K^{(0)}-\IZ_K^{(1)}$ symmetry. The existence of these 't Hooft anomalies restrict the consistent gaugings:
\begin{itemize}
\item[1.] 0-form axion shift $\mathbb{Z}_K^{(0)}$ or its subgroup $\IZ_n^{(0)} \subset \mathbb{Z}_K^{(0)}$: 

This is what we have focused so far. From the first term in the inflow action eq.~(\ref{eq:subgroup_gauging_inflow_action}) we see that such a gauging turns some of the 't Hooft anomalies among global symmetries into ABJ-type anomalies between background gauge fields of global symmetry and dynamical gauge fields, hence resulting in quantum mechanical breaking of global symmetries. It leads to the modification of $U(1)^{(2)}$ and breaks $\mathbb{Z}_K^{(1)} \to \mathbb{Z}_{K/n}^{(1)}$. This in turn leads to changes of line and surface (cosmic string) operators of the theory.
\item[2.] 1-form electric $\mathbb{Z}_K^{(1)}$ or its subgroup $\IZ_n^{(1)} \subset \mathbb{Z}_K^{(1)}$: 

Due to the 3-group structure, gauging $\mathbb{Z}_K^{(1)}$ or its subgroup alone is not a consistent operation. Rather, the 3-group transformation rules eq.~(\ref{eq:axion_MW_Am3_gauge_transf}) imply that gauging $\IZ_n^{(1)} \subset \mathbb{Z}_K^{(1)}$ ($n$ may be equal to $K$) must be accompanied by gauging of $\IZ_n^{(2)} \subset U(1)^{(2)}$. The anomalies then imply that $\IZ_K^{(0)}$  has an ABJ anomaly that completely breaks and $U(1)^{(1)}$ is modified so that it has periodicity $2\pi n$.\footnote{Although $\IZ_K^{(0)}$ is broken as a global symmetry, it participates in a non-invertible symmetry structure. See \cite{Choi:2022jqy, Cordova:2022ieu, Choi:2022fgx} for related discussions.  }  

\item[3.] 2-form axion winding $U(1)^{(2)}$ or its subgroup $\IZ_n^{(2)} \subset U(1)^{(2)}$:

Here the anomaly becomes an ABJ anomaly for $\IZ_K^{(0)}$ that breaks the symmetry completely while leaving the other parts of the 3-group untouched. 
%
\item[4.] 1-form magnetic $U(1)^{(1)}$ or its subgroup $\mathbb{Z}_n^{(1)} \subset U(1)^{(1)}$:

Here the anomaly becomes an ABJ anomaly for $\IZ_K^{(1)}$ that breaks the symmetry completely while leaving the other parts of the 3-group untouched. 

\end{itemize} 

\noindent It would be interesting to analyze each of these cases  and understand the observable consequences of these discrete gaugings (e.g.~local, line, and surface operator spectrum). Furthermore, it is an interesting question which of these TQFT couplings can arise as a long-distance effective description of more fundamental QFTs at short distance scales. We leave this analysis to future investigations.

\section{Brief Comments on Phenomenological Implications}
\label{sec:pheno}

We conclude this paper by briefly commenting on the potential phenomenological implications of the scenarios discussed in this paper. Further studies are certainly required make precise predictions, which we leave for a future work.  Our discussion here will be brief and qualitative, highlighting the potential difference with well studied signals of the cosmic strings (see \cite{Vachaspati:2015cma} for a review). 

We begin with the discussion of the implications which are largely independent of the UV completions. 
The main focus of this paper is a Higgsed $U(1)_B$ gauge symmetry as the origin of a TQFT.  However, there is another unbroken $U(1)_A$ gauge symmetry in the story. To be more concrete, we will proceed by first assuming it is the SM $U(1)_{\rm{EM}}$.  A TQFT does not have low energy excitations, hence it is ``absent'' in the IR. Yet, it still leaves some imprints.

As described in detail in this paper, a main portal can be an axion with coupling to a TQFT, as shown in eq.~\eqref{eq:S_1}.  Such a coupling does lead to important differences in comparison with the ``usual'' well studied axion strings. Potential signals for axion strings with localized fermionic zero modes have been studied \cite{Naculich:1987ci,Kaplan:1987kh,Agrawal:2020euj,Witten:1984eb,Manohar:1988gv,Fukuda:2020kym}. However, the emphasis has  been on the fermion which is charged under the SM $U(1)_{\rm{EM}}$ in which case axion strings will be charged up by passing through regions of magnetic field in the universe. In our case, the fermionic zero modes localized on the string are required to carry specific $U(1)_B$ charges in addition to the $U(1)_A$ charge. This leads to some remarkable differences. 

First of all, since $U(1)_B$ is Higgsed, there are no macroscopic regions with non-zero $U(1)_B$ field. However, if we assume the unbroken $U(1)_A$ is the SM $U(1)_{\rm{EM}}$, then the axion strings will be similarly charged up by the magnetic fields in the universe. One of the interesting consequences of the $U(1)_B$ charges carried by the zero mode fermions is that they can potentially affect the fate of the string loops. 

As pointed out in \cite{Carter:1993wu,Brandenberger:1996zp,Martins:1998gb,Martins:1998th,Carter:1999an,Fukuda:2020kym,Ibe:2021ctf,Abe:2022rrh}, charged fermions present on the string can provide a pressure which prevent the string from shrinking, leading to potentially stable final string loops (vorton). However, fermions with EM charges can be expected to decay into SM charged particles which leads to the decay of the vorton \cite{Fukuda:2020kym,Ibe:2021ctf}. Such a decay, however, would not be possible in our case with the absence of the light particles charged under $U(1)_B$. While the vorton may still be able to lose charge by scattering processes~\cite{Agrawal:2020euj}, this could lead to changes in the evolution of the string loops and additional phenomenological consequences.
We emphasize that,  from the point of view of the IR theory, this could be attributed to a global symmetry which is imposed ``by hand''. However, in the scenario discussed here, we see that it is a consequence of the TQFT coupling and requirement of anomaly matching. 
It is  possible that the $U(1)_{\rm{PQ}}$ is also broken explicitly, such as by the instanton effect in the QCD. In this case, a string domain-wall network will form and collapse. The anomaly on the string then implies a certain Chern-Simons theory living on the domain wall \cite{Gaiotto:2017tne,Hidaka:2020iaz}. Additionally, it has been shown that a charged axion string can influence the evolution of the string network and the dynamics of string domain-wall network \cite{Fukuda:2020kym}. It would be interesting to investigate further such phenomena in our case. 

As discussed in Ref.~\cite{Agrawal:2020euj}, it is expected that axion string is approximated electrically neutral due to the Schwinger pair production of SM light charged particles in the vicinity of the string, in the strong electromagnetic field produced by the charged particles localized on the string. Such an effect would not be effective for the BF charge on the string, since the BF-charged fermions are heavy and the $U(1)_B$ is Higgsed outside of the string . Hence, if the axion string is charged up with BF-charged particles, it would not be neutral in the BF charge. At the same time, we do not expect this would lead to macroscopic effect since the $U(1)_B$ field is short ranged.

If BF strings are present in the universe, they can in principle lead to different signals as well. It is generically expected that their tension will be different from the axion string. A more unique feature of the BF string is the non-trivial holonomy of the $U(1)_B$ gauge field around the string. In principle, BF-charged particles (for example DM candidate), passing around the BF string could experience Aharonov-Bohm effect which changes their distributions. It remains to be investigated whether this can lead to observable effects.

Thus far  we have discussed the possible IR (universal) signals. However, signals that depend on UV completion can be equally important. In addition to enhancing the discovery potential, they also provide complementary information which may eventually lead to a more complete picture. We will briefly mention a couple such possibilities in the following.

The mechanisms discussed in this paper are largely independent of the absolute scales of UV symmetry breaking. At the same time, the potential signals are sensitive to the scales. A large class of signatures of the cosmic strings are through their gravitational interactions, which is highly sensitive to the string tension $\mu$,  with the current limit roughly in the range of $G\mu < 10^{-8}$ to $10^{-9}$ \cite{Danos:2008fq,Blanco-Pillado:2013qja,Jenet:2006sv} and may potential reach $G\mu < 10^{-10} $ to $10^{-12}$  in the near future \cite{Khatri:2008zw}. If the cosmic string discussed here can give rise to such signals, it will certainly provide highly valuable information.   This is especially important for the second case of TQFT coupling discussed in the paper, in which the main feature is the varying tension of the string spectrum. 

If both the axion string and the BF string are present in the universe, there is potential for richer dynamics. In particular, the existence of two different kinds of strings differs from well studied scenarios. For example, if the symmetry breaking dynamics are such that it is energetically favored for the strings to overlap, they would tend to align rather than cut through each other. This allows for the   possibility of producing  co-axial strings with different properties and can also potentially affect the evolution of the string network. There can also be interesting differences with the standard axion string story depending on the relative scale of the $U(1)_B$ and $U(1)_{\rm{\rm PQ}}$ breaking. Either one can happen at higher scale in the UV theory for the first type of TQFT coupling. In particular, if $U(1)_{\rm{\rm PQ}}$ breaks at a higher scale (earlier in the evolution of the universe), there could be an epoch in which the universe is filled with a background of primordial (unbroken) $U(1)_B$ field. This can charge up the axion string through the interaction with the zero mode fermions. The influence of the primordial electromagnetic field on the evolution of the axion string network has been studied \cite{Fukuda:2020kym}. It would be interesting to generalize it to the case of a primordial $U(1)_B$ background field. 

We have been assuming that the unbroken $U(1)_A$ is the SM $U(1)_{\rm{EM}}$. However, $U(1)_A$ could instead be a dark photon. Another intriguing  possibility is that the gauge boson of the $U(1)_B$ is the dark photon. Instead of (or in addition to) coupling to the SM model via a kinetic mixing with the photon, it couples through the TQFT portal described in this paper. Since it is common to assume the dark photon mass is small, this could be an extreme example of the case in which PQ symmetry breaking happens at a much higher scale. It has been pointed out recently \cite{East:2022rsi} that cosmic string associated with dark photon can be produced in a broad range of dark photon production scenarios, such as through the axion coupling in eq.~\eqref{eq:S_0}. Hence, this would be a natural stage to study the interplay between these two kinds of strings and implication of the couplings in eq.~\eqref{eq:S_1}.

\section*{Acknowledgments}

We are grateful to Clay C\'{o}rdova, Thomas Dumitrescu, Jeffrey Harvey, Seth Koren, Shu-Heng Shao for helpful discussions and related collaborations.
S.H.\ was supported by the DOE grants DE-SC-0013642 and  DE-AC02-06CH11357.  LTW is supported
by the DOE grant DE-SC0013642. TDB is supported by Simons Foundation award 568420 (Simons Investigator) and award 888994 (The Simons Collaboration on Global Categorical Symmetries).

\appendix

\section{Brief Introduction to Generalized Global Symmetries}
\label{app:review_GGS}

In this appendix, we provide a brief introduction to generalized global symmetries, hopefully in an easy and accessible way.
The modern notion of global symmetry, in addition to group-like symmetries, also includes non-group-like symmetries such as higher-group symmetry, non-invertible symmetries, and subsystem symmetries. In this section, we mainly focus on higher-form symmetries.  We refer to \cite{Cordova:2022ruw} and references therein for more extensive and detailed discussion.

\subsection{Ordinary symmetry}
\label{app:0-form symm}

We begin our discussion by recalling the general properties of ordinary (i.e. 0-form) symmetries in the language that admits straightforward generalizations to higher-form symmetries. According to Noether's theorem, an ordinary continuous symmetry corresponds to a conserved current 
\beq
\partial_\mu j_1^\mu = 0
\eeq  
In differential-form notation, the conserved current is written in terms of a co-closed 1-form:
\beq
\label{coclosed}
d * j_1 = 0~.
\eeq 
Here $* j_1$ is a 3-form that is the Hodge dual of $j_1$. By definition, the existence of  a symmetry means there are charged objects which transform under the symmetry. In the case of the ordinary (0-form) symmetry, 
charged objects are local (i.e.~0-dimensional, hence the name 0-form symmetry) operators $\mathcal{O} (x)$. For example, they can be ~elementary or composite field operators. In general, $\mathcal{O} (x)$ transforms under a symmetry transformation $g$ as
\beq
  \mathcal{O} (x) \mapsto R(g)\cdot \mathcal{O} (x)
\eeq
where $R(g)$ denotes the representation of $g$. A set of ordinary symmetry transformations $\{ g \}$ often forms a group $G$, which can be either continuous or discrete. The group $G$ can be either abelian or non-abelian. More recently,  it has become clearer that there exist symmetries whose mathematical structures are not  groups such as higher-groups \cite{Cordova:2018cvg, Benini:2018reh, Brennan:2020ehu, Hidaka:2020iaz, Hidaka:2020izy}  (as we also discuss in Section~\ref{subsec:3-group_discrete_gauging} and Appendix~\ref{subapp:TQFT-coupling I_3-group}) and non-invertible \cite{Kaidi:2021xfk, Choi:2021kmx, Cordova:2022ieu, Choi:2022jqy, Cordova:2022fhg,Roumpedakis:2022aik, Choi:2022zal, Choi:2022rfe, Choi:2022fgx, Bhardwaj:2022yxj, Arias-Tamargo:2022nlf, Bhardwaj:2022lsg, Bartsch:2022mpm, GarciaEtxebarria:2022vzq, Kaidi:2022cpf,Yokokura:2022alv, Bhardwaj:2022kot, Bhardwaj:2022maz, Bartsch:2022ytj} 
 symmetries which strictly speaking are described by category theory.\footnote{For applications of higher-group and non-invertible symmetries in particle physics model building, see \cite{Cordova:2022fhg, Cordova:2022qtz}.} 

It is often useful to analyze the symmetry properties of a theory by coupling it to background gauge fields. For a continuous ordinary global symmetries, we can couple 
the conserved current to a background gauge field $A_\mu$
\beq
Z \left[ A \right] = \int \left[ d \psi \right] ~{\rm exp}\left\{i\, S \left[ \psi \right] + i \int d^d x\, A_\mu (x) j_1^\mu (x)\right\}~.
\eeq
Here, we used $\psi$ to denote collectively all quantum fields of the theory.\footnote{We emphasize that the theory under consideration needs not have a Lagrangian description. We simply imagine having an action here to streamline the discussion.}

In terms of differential forms, we can couple a 0-form continuous symmetry to a background gauge field $A^{(1)}$ as
\eq{
Z[A^{(1)}]=\int [d\psi]~{\rm exp}\left\{i\, S \left[ \psi \right] + i \int A^{(1)} \wedge \ast j_1 \right\},
}
Since a 0-form symmetry has a conserved current that is a 1-form (its dual $*j_1$ is $(d-1)$-form in $d$ spacetime dimensions), the background gauge field $A^{(1)}$ is a 1-form gauge field. 
The conservation of $j_1$ \eqref{coclosed} then implies that the above partition function is invariant under background gauge transformations $A^{(1)} \to A^{(1)} + d \lambda_0$, where $\lambda_0$ is a 0-form transformation parameter.\footnote{Here, we imagine an abelian symmetry such as $U(1)$ for simplicity sake. It can be easily generalized to non-abelian cases for which the background gauge transformation takes the familiar form $\delta_\lambda A_\mu = \left[ \lambda, A_\mu \right] + \partial_\mu \lambda$. Since higher-form group symmetries are abelian, 
it is sufficient to focus on the $U(1)$ case.} 

In a symmetry preserving vacuum, all non-vanishing expectation values $\langle \mathcal{O} (x) \cdots \rangle$ are invariant under $G$ transformations. The invariance under infinitesimal $G$ transformations implies the Ward identity
\beq
\partial_\mu j_1^\mu (x) \mathcal{O} (y) = \delta^{(d)} (x-y) \,R(Q)\cdot \mathcal{O} (y)
\label{eq:0form_Ward_infin}
\eeq
where $R (Q)$ is the generator for the infinitesimal transformation in the representation $R$.  

Using the Ward identity, one can then define a charge operator that generates the $G$ symmetry in the quantum theory by integrating the dual current $*j_1$ over a closed $(d-1)$-manifold $\Sigma_{d-1}$
\beq
Q (\Sigma_{d-1} ) = \int_{\Sigma_{d-1}} *j_1~.
\label{eq:charge}
\eeq
This charge operator $Q( \Sigma_{d-1})$ is topological in a sense that any correlation function containing $Q(\Sigma_{d-1})$ is unchanged under  continuous deformations of the manifold $\Sigma_{d-1}$ so long as such a deformation does not cross a charged operator.

In the more familiar case of $d=4$, we often choose $\Sigma_3$ to be a spatial slice at a fixed time on which the Hilbert space is defined, and the above expression becomes
\beq
Q (\Sigma_3) = \int d^3 x \; j^0~.
\eeq
However, in general we can choose any closed $(d-1)$-manifold to define the charge operator due to the topological nature of the definition in eq.~(\ref{eq:charge}).  %

This allows us to define  a topological operator  for any group element $g=e^{i\lambda}$ by exponentiating the charge operator
\beq
U (g, \Sigma_{d-1}) = {\rm exp} \left( i \lambda 
 Q (\Sigma_{d-1}) \right)
\label{eq:0form_SDO}
\eeq
called a \emph{symmetry defect operator}. 
This operator is topological in the same way as  $Q(\Sigma_3)$.  More explicitly, suppose $\Sigma_{d-1}^\prime$ is a small continuous deformation of $\Sigma_{d-1}$ without crossing any charged local operators, then the difference of the symmetry defect operator wrapped on the two different manifolds is given by %
\eq{
U(g, \Sigma)\cdot U(g,\Sigma^\prime)^{-1}&=U(g,\Sigma) \cdot U (g^{-1}, \Sigma^\prime ) = {\rm exp} \left( i \lambda \left( \int_{\Sigma} *j_1 - \int_{\Sigma^\prime} *j_1 \right) \right)\\
& = {\rm exp} \left( i \lambda \int_{\bar{\Sigma}} d *j_1 \right) = 1
}
where $\bar{\Sigma}$ is 4 dimensional manifold whose boundary is the union of $\Sigma$ and $\Sigma^\prime$ (where $\Sigma^\prime$ has  opposite orientation). Here we used the fact that $g^{-1}=e^{-i\lambda}$ and the conservation equation $d*j_1=0$. 

Using similar manipulations, it is easy to show that the symmetry defect operators satisfy the $G$  multiplication law%
\beq
U (g_1, \Sigma_{d-1})\cdot U (g_2, \Sigma_{d-1}) = U (g_3, \Sigma_{d-1})
\label{eq:GSS_SDO_group_multiplication}
\eeq
with $g_3 = g_1 g_2$. 

The Ward identity eq.~(\ref{eq:0form_Ward_infin}) %
then implies that these symmetry defect operators (SDO) implement the $G$ action on charged operators that cross its world volume. For example, if we consider $\Sigma_{d-1}$ that links the point $x$, then wrapping a SDO for the element $g$ on $\Sigma_{d-1}$ will act on a charged operator $\CO(x)$ as\footnote{A $(d-1)$-manifold $\Sigma_{d-1}$ that wraps a point $x$ is one that can be contracted to a point by passing through $x$. The relation \eqref{transformationproperty} then follows from the topological property of $U(g,\Sigma_{d-1})$ by contracting it to a point (the trivial operator) and acting on $\CO(x)$ as $\Sigma_{d-1}$ passes through $x$. 
}

\beq
\label{transformationproperty}
U (g, \Sigma_{d-1})\, \mathcal{O} (x) = R (g)\cdot \mathcal{O} (x)~. 
\eeq

We can similarly define symmetry defect operators for discrete 0-form global symmetries. These are again topological operators that implement symmetry transformations on charged local operators. This structure of a global symmetry corresponding to the existence of topological operators can be taken as a definition or, if the reader prefers, we can think of discrete abelian symmetries as being part of some continuous abelian symmetry which is broken to a discrete subgroup (explicit or otherwise), in which case the structure of symmetry defect operators is inherited from the continuous completion.

\subsection{Higher-form symmetry}
\label{app:higher-form symm}

We can generalize the discussion in the previous subsection to higher-form symmetries. 
A $p$-form global symmetry in a QFT defined on $d$-dimensional spacetime, denoted $G^{(p)}$, acts on charged objects supported on $p$-dimensional manifolds (obviously $p \leq d$). For continuous $p$-form symmetry, it has $p+1$-form conserved current
\beq
d * j_{p+1} = 0~.
\eeq
When there is a current one can construct the charge and symmetry defect operators as before
\beq
Q (\Sigma_{d-p-1}) = \int_{\Sigma_{d-p-1}} * j_{p+1}~, \;\;\;\;\; U (g, \Sigma_{d-p-1}) = {\rm exp} \left( i \lambda \oint_{\Sigma_{d-p-1}} *j_{p+1} \right)
\eeq
where now the charge operator and symmetry defect operators are defined on codimension $p+1$ manifolds.\footnote{For $d$ space-time dimensional space, a codimension $p+1$ manifold has $d-p-1$ dimensions. }

The conservation of $j_{p+1}$ ensures that the associated symmetry defect operator is topological -- the argument follows analogously to the case of 0-form symmetry described in Section~\ref{app:0-form symm}.
If the collection of symmetry transformations $\{ g \}$ forms a group $G$, the products of symmetry defect operators furnish the group multiplication law. For any $p>0$, a $p$-form symmetry group is necessarily abelian -- this follows from the fact that there is no topologically invariant ordering of co-dimension $p+1 \geq 2$ manifolds to accommodate non-abelian multiplication.\footnote{The reason is that any pair of marked manifolds $\Sigma^{(1)},\Sigma^{(2)}$ of codimension $\geq 2$ can be freely deformed $\Sigma^{(1)},\Sigma^{(2)}\longmapsto \widetilde\Sigma^{(1)},\widetilde\Sigma^{(2)}$ so that $\Sigma^{(1)}\cong \widetilde\Sigma^{(2)}$ and $\Sigma^{(2)}\cong \widetilde \Sigma^{(1)}$. For example, any two loops in $\IR^3$ can be exchanged by smooth deformations. }

A $p$-form symmetry acts on $p$-dimensional operators $W_p (m, \Sigma_p)$ where we take the notation that $m$ is the charge. 
Again, there is an associated Ward identity, which leads to an action of the symmetry defect operator on $W_p (m, \Sigma_p)$
\beq
U (g, \Sigma_{d-p-1}) W_p (m, \Sigma_p) = {\rm exp} \Big( i \lambda m\text{ Link}( \Sigma_{d-p-1}, \Sigma_p ) \Big) W_p (m, \Sigma_p)~,\quad g=e^{i\lambda}
\eeq
similar to the case of the 0-form symmetry, 
where $\text{ Link}( \Sigma_{d-p-1}, \Sigma_p )$ is the linking number of the two manifolds (see \cite{Horowitz:1989km} for an explanation of linking number in the language of QFT). This equation holds for both continuous (e.g. $\lambda \in \mathbb{S}^1$ for $U(1)$) and discrete (e.g.~$\lambda = 2\pi / n$ for $\mathbb{Z}_n$) symmetries.

In the rest of this section, we will demonstrate these points with a simple and tractable example with continuous global symmetries. We will discuss an analogous example with  discrete higher-form symmetry ($\mathbb{Z}_n$ gauge theory) in Appendix~\ref{app:review_BF}.

Consider Maxwell theory in $(3+1)$ dimension. This theory enjoys $U(1)_e^{(1)}$ 1-form electric symmetry and $U(1)_m^{(1)}$ 1-form magnetic symmetry. The action is written
\beq
S = \frac{1}{2g^2} \int F^{(2)} \wedge * F^{(2)}
\eeq
where $F^{(2)} = d A^{(1)}$ is the field strength of 1-form gauge field $A^{(1)}$. The equation of motion and Bianchi identity are written as
\eq{
d * F^{(2)} = 0~,\quad \quad 
d F^{(2)} = 0~. \label{eq:MW_Bianchi}
}
These equations can be thought of as current conservation equations, where the former is interpreted as the conservation of a 1-form electric symmetry and the latter is interpreted as the conservation of a dual magnetic 1-form symmetry. Note that for $d=4$, both $F^{(2)}$ and its dual $* F^{(2)}$ are 2-forms, and these are 2-form currents for a pair of dual 1-form global symmetries. 

The 1-form electric symmetry $U(1)_e^{(1)}$ has a 2-form current and symmetry defect operators
\beq
J_2^e = \frac{i}{g^2} F^{(2)}~,\quad U_e (g, \Sigma_2) = {\rm exp} \left( \frac{i \lambda}{g^2} \oint_{\Sigma_2} * F^{(2)} \right)~.
\label{eq:1form_MW_electric_current_SDO}
\eeq
for $g=e^{i \lambda}\in U(1)$. This symmetry is an ``electric'' symmetry because $\oint_{\Sigma_2} * F^{(2)}$ measures the electric flux through $\Sigma_2$ and it acts on the dynamical electric gauge field by a shift $A^{(1)} \to A^{(1)} + \lambda_e^{(1)}$ where the transformation parameter $\lambda_e^{(1)}$ itself is a closed 1-form (i.e.~a flat gauge connection)  normalized as $\oint \lambda_e^{(1)} \in U(1)$.\footnote{The relation of the 1-form parameter $\lambda_e^{(1)}$ and $\lambda \in 2\pi \mathbb{Z}$ appearing in eq.~(\ref{eq:1form_MW_electric_current_SDO}) is the following.
\beq
\oint_{M_4} d \lambda_e^{(1)} \wedge * F^{(2)} = \lambda \oint_{\Sigma_2} * F^{(2)}~.
\eeq
}

The gauge invariant operators that are charged under $U(1)_e^{(1)}$ are the Wilson line operators $W_1 (m, \Sigma_1) = e^{i m \int_{\Sigma_1} A^{(1)}}$, where $ m \in \mathbb{Z}$. It is acted on by a non-trivial linking with the symmetry defect operator
\beq
U_e (g, \Sigma_2) W_1 (m, \Sigma_1) = {\rm exp} \left( i \lambda m \text{ Link}(\Sigma_2, \Sigma_1) \right) W_1 (m, \Sigma_1)~,\quad g=e^{i \lambda}
\label{eq:1form_MW_Electric_Ward_id}
\eeq 
where $\text{Link}(\Sigma_2, \Sigma_1)$ denotes the linking number between $\Sigma_2$ and $\Sigma_1$. 

The 1-form electric symmetry can be explicitly broken by introducing electrically charged fields. The presence of those electrically charged particles modifies the equation of motion to 
\beq
d* J_2^e = \frac{i}{g^2} d * F^{(2)} = j^{(3)}_{\rm charge}
\label{eq:ChargeBreak1formE}
\eeq
which violates the conservation law for $J_2^e$. Here, $j^{(3)}_{\rm charge}$ is the Hodge dual of the ``usual'' 1-form momentum density for the charged particles. If the particles have charge $n$, the source in eq. \eqref{eq:ChargeBreak1formE} breaks $U(1)_e^{(1)} \to \mathbb{Z}_n^{(1)}$. Physically, this breaking occurs because a dynamical field with charge $n$ can pair produce to break $n$ Wilson lines whose charge is a multiple of $n$. This means that the Wilson line charge is only preserved mod $n$ and consequently $U(1)^{(1)}_e$ is broken $U(1)^{(1)}_e\to \IZ_n^{(1)}$.  

The 1-form magnetic symmetry $U(1)^{(1)}_m$ has a 2-form current and associated symmetry defect operator
\beq
J_2^m = \frac{1}{2\pi} * F^{(2)}~,\quad  U_m (g, \Sigma_2) = {\rm exp} \left( \frac{i \lambda}{2\pi} \oint_{\Sigma_2} F^{(2)} \right)
\label{eq:1form_MW_magnetic_current_SDO}
\eeq
for $g=e^{i\lambda}\in U(1)$. We say that this symmetry is ``magnetic'' because $\oint_{\Sigma_2}  F^{(2)}$ measures the magnetic flux through $\Sigma_2$ and consequently this symmetry acts on the dual magnetic photon $\tilde{A}^{(1)}$ by a shift $\tilde{A}^{(1)} \to \tilde{A}^{(1)} + \lambda_m^{(1)}$ where $\lambda_m^{(1)}$ is a closed 1-form (i.e. a flat gauge connection) normalized as $\oint\lambda_m^{(1)}\in U(1)$. 

The gauge invariant operators that are charged under $U(1)^{(1)}_m$ 
 are 't Hooft line operators $T_1 (\ell, \Sigma_1) = e^{i \ell \oint_{\Sigma_1} \tilde{A}^{(1)}}, \; \ell \in \mathbb{Z}$:
\beq
U_m (g, \Sigma_2) T_1 (\ell, \Sigma_1) = {\rm exp} \left( i \lambda \ell\text{ Link} (\Sigma_2, \Sigma_1) \right) T_1 (\ell, \Sigma_1)~,\qquad g=e^{i \lambda}~.
\label{eq:1form_MW_Magnetic_Ward_id}
\eeq 
The magnetic 1-form global symmetry can be broken by dynamical monopoles, similar  to the way the 1-form electric symmetry is broken by electrically charged particles. Such dynamical states modify the Bianchi identity to 
\beq
\frac{1}{2\pi} d F^{(2)} = j_{\rm mon}^{(3)}~. 
\eeq
As with the Wilson lines and charged particles, dynamical monopoles can break 't Hooft lines by monopole-anti-monopole pair production, thereby breaking $U(1)^{(1)}_m$. 

We can couple the Maxwell theory to background gauge fields of these two 1-form global symmetries. The action with such couplings are given by
\beq
S = \frac{1}{2g^2} \int \left( F^{(2)} - \mathcal{B}_e^{(2)} \right) \wedge * \left( F^{(2)} - \mathcal{B}_e^{(2)} \right) + \frac{i}{2\pi} \int \mathcal{B}_m^{(2)} \wedge \left( F^{(2)} - \mathcal{B}_e^{(2)} \right)
\label{eq:MW with BGFs}
\eeq 
where $\mathcal{B}_e^{(2)}$ and $\mathcal{B}_m^{(2)}$ are the 2-form background gauge fields of the electric and magnetic 1-form symmetries. They transform under the respective symmetry as background gauge transformations.
\beq
\mathcal{B}_e^{(2)} \to \mathcal{B}_e^{(2)} + d \lambda_e^{(1)}~,\quad  \mathcal{B}_m^{(2)} \to \mathcal{B}_m^{(2)} + d \lambda_m^{(1)}~.
\eeq
Note that in addition to coupling $\CB_{e,m}^{(2)}$ to their respective currents -- analogous couplings $i \int A^{(1)} \wedge * j_1$ -- we have also added additional background counterterms to make the theory explicitly invariant under $U(1)_e^{(1)}$ background gauge transformations which shifts  $F^{(2)} \to F^{(2)} + d \lambda_e^{(1)}$.

However, now we see that the action is not invariant under  $U(1)_m^{(1)}$ gauge transformations
\eq{
\delta S=-\frac{i}{2\pi}\int \lambda_m^{(1)}\wedge d\CB_e^{(2)}~.
} 
We can make a different choice for local counterterms that makes the action invariant under $U(1)_m^{(1)}$ background gauge transformations so that the theory is given by 
\eq{
S^\prime = \frac{1}{2g^2} \int \left( F^{(2)} - \mathcal{B}_e^{(2)} \right) \wedge * \left( F^{(2)} - \mathcal{B}_e^{(2)} \right) + \frac{i}{2\pi} \int \mathcal{B}_m^{(2)} \wedge F^{(2)} 
\label{MW with alt ct}
}
However, we now see that the theory is not invariant under $U(1)_e^{(1)}$ gauge transformations
\eq{
\delta S^\prime=-\frac{i}{2\pi}\int \lambda_e^{(1)}\wedge d\CB_m^{(2)}~.
}
In fact, one can show that no choice of local counterterms makes the theory invariant under both electric and magnetic 1-form symmetries with generic background gauge fields $\mathcal{B}_e^{(2)}$ and $\mathcal{B}_m^{(2)}$ turned on. This means that we can not gauge both global symmetries. Such a ``tension'' among different global symmetries is indicative of a 't Hooft anomaly involving both $U(1)_e^{(1)},U(1)_m^{(1)}$. 
  
  It is often useful to organize anomalies of $d$ dimensional quantum field theory in terms of a  $d+1$ dimensional topological quantum field theory. This is known as anomaly inflow \cite{Callan:1984sa, Witten:2019bou} (see also \cite{Hong:2020bvq} for a recent discussion of anomaly inflow in the context of AdS/CFT duality and with relevance to particle phenomenology). In our current example, the $U(1)_e^{(1)}U(1)_m^{(1)}$ mixed anomaly is described by a 5d anomaly TQFT:
\beq
S_{\rm inflow} = - \frac{i}{2\pi} \int_{N_5} \mathcal{B}_m^{(2)} \wedge d \mathcal{B}_e^{(2)}
\eeq
where $\partial N_5 = M_4$, i.e.~the boundary of the auxiliary 5d manifold $N_5$ is the 4d spacetime. In fact, one easily sees that under a magnetic transformation $\mathcal{B}_m^{(2)} \to \mathcal{B}_m^{(2)} + d \lambda_m^{(1)}$ with $\mathcal{B}_e^{(2)}$ activated (or similarly under an electric transformation with $\mathcal{B}_m^{(2)}$ turned on) this reproduces the same anomaly as the original $4d$ action eq.~(\ref{eq:MW with BGFs}).

\section{$\mathbb{Z}_n$ TQFT}
\label{app:review_BF}

In this appendix, we review an example of a topological field theory, also known as a BF theory, that is a $\IZ_n$ gauge theory which was introduced in \cite{Horowitz:1989ng, Horowitz:1989km}. This theory appears ubiquitously in the literature on generalized global symmetry (see \cite{Banks:2010zn, Kapustin:2014gua, Gaiotto:2014kfa,Maldacena:2001ss,Alford:1991vr} for a useful introduction).\footnote{Among other things, BF theory are prototypical discrete gauge theories and can be used to describe the IR theory of spontaneously broken discrete higher-form symmetries \cite{Anber:2021iip}. 
} 

The action for this $4d$ $\mathbb{Z}_n$ TQFT is given by\footnote{Although our discussion here is focused on $4d$, most of the details generalize straightforwardly to any dimension. }%
\beq
S_{\rm BF} = \frac{i n}{2\pi} \int B^{(2)} \wedge d A^{(1)}~.
\label{eq:app_BF_action}
\eeq
The integrand can also be written as $B^{(2)} \wedge F^{(2)}$, hence the name BF theory.
While this theory has many subtleties associated to the fact that it describes a discrete gauge theory, it is illuminating to keep in mind a particularly simple UV completion. 

Consider an Abelian Higgs model with a charge $n$ Higgs field $\Phi$.
\beq
\mathcal{L} = \left| d \Phi - i n A^{(1)} \Phi \right|^2+ \frac{1}{2g^2} F^{(2)} \wedge * F^{(2)} - V (\Phi)~.
\label{eq:AHM_linear}
\eeq
where $V(\Phi)$ is chosen so that $\Phi$ condenses in the the IR. It is clear that the condensation of $\Phi$ will break the $U(1)$ gauge group down to $\mathbb{Z}_n$ since the vev of $\Phi$ is invariant under  $\mathbb{Z}_n\subset U(1)$ gauge transformations. Thus, in the IR the theory will flow to a $\mathbb{Z}_n$ gauge theory. 

To see this, note that the radial mode of  $\Phi$ has a mass similar to the scale of the symmetry breaking and is integrated out. Hence, we can decompose $\Phi = \Lambda e^{i \varphi}$ where $\Lambda$ is the scale of the symmetry breaking and $\varphi$ is a periodic scalar field of charge $n$. Substituting this into the above action yields
\beq
\mathcal{L} \sim \Lambda^2 \left( d \varphi - n A^{(1)} \right) \wedge * \left( d \varphi - n A^{(1)} \right) + \frac{1}{2g^2} F^{(2)} \wedge * F^{(2)}~.
\label{eq:AHM_nonlinear}
\eeq
In the low energy limit, which effectively sends $\Lambda \to \infty$, the gauge field is fixed $A^{(1)}= \frac{d \varphi}{n}$ (i.e.~pure gauge configuration) setting the gauge kinetic term to zero -- leaving no local degrees of freedom. This is not a surprising statement at all since all but discrete part of scalar degrees of freedom are ``eaten'' by the gauge boson. The key point however, is that there is still an important discrete remnant. 

Let us study the action \eqref{eq:AHM_nonlinear} a bit more closely. 
To proceed further, it is useful to dualize $\varphi$. This can be achieved by introducing a new 3-form field $H$ and rewriting the Lagrangian 
as
\beq
\mathcal{L} = \frac{1}{(4 \pi)^2 \Lambda^2} H \wedge * H + \frac{i}{2 \pi} H \wedge (d \varphi - n A^{(1)} )~.
\eeq
One can check that by integrating out $H$ using its equation of motion, $*H=4 \pi i \Lambda^2 (d \varphi - n A^{(1)})$, we recover the Lagrangian in eq.~(\ref{eq:AHM_nonlinear}). Additionally, the equation of motion for $\varphi$, $d H = 0$, which means we can locally introduce a 2-form  $B^{(2)}$ with $d B^{(2)} =H$. The Lagrangian then becomes
\beq
\mathcal{L} =  \frac{1}{(4 \pi)^2 \Lambda^2} H \wedge * H + \frac{in}{2 \pi} B^{(2)} \wedge d A^{(1)}~.
\eeq
Taking the limit $\Lambda \to \infty$, we arrive at the BF theory in eq.~(\ref{eq:app_BF_action}).

Having shown that eq.~(\ref{eq:app_BF_action}) describes a $\mathbb{Z}_n$ gauge theory, we now study it in more detail. First note that the theory is completely independent of metric and hence is a ``topological'' field theory.\footnote{In the language of differential forms, the metric only enters via the Hodge star operation.} 
Additionally, the equations of motion set the two field strengths to vanish
\beq
d A^{(1)} = d B^{(2)} = 0~.
\eeq
This eliminates all local degrees of freedom in the IR, and confirms once again that the theory is topological. 

This theory also has two $U(1)$ gauge symmetries
\bea
A^{(1)}  \to  A^{(1)} + d \lambda^{(0)} ~, \qquad
B^{(2)}  \to  B^{(2)} + d \lambda^{(1)}~.
\eea
The $A^{(1)}$ gauge symmetry follows from the UV theory while the $B^{(2)}$ gauge symmetry is a consequence of the fact that $B^{(2)}$ is dual to $\varphi $.\footnote{
More explicitly, the vortices of $\varphi$ correspond to the Wilson-like surface operators of $B^{(2)}$. The fact that $\varphi$ is periodic means that these vortices are integer quantized which requires that $B^{(2)}$ is a 2-form gauge field with the transformation properties above.  
} 
These two gauge fields have corresponding 
$\mathbb{Z}_n$ higher-form global symmetries under which fields transform as
\eq{
& \mathbb{Z}_n^{(1)} : \;\;  A^{(1)} \to  A^{(1)} + \frac{1}{n} \epsilon^{(1)}~, \qquad \oint \epsilon^{(1)} = 2\pi \mathbb{Z} \\
& \mathbb{Z}_n^{(2)} : \;\;  B^{(2)} \to  B^{(2)} + \frac{1}{n} \epsilon^{(2)}~, \qquad\oint \epsilon^{(2)} = 2\pi \mathbb{Z}.
}
The existence of these symmetries is also manifested by the presence of gauge invariant Wilson line and surface operators:
\beq
W_1 (\ell, \Sigma_1) = {\rm exp} \left( i \ell \oint_{\Sigma_1} A^{(1)} \right)~,\quad W_2 (m, \Sigma_2) = {\rm exp} \left( i m \oint_{\Sigma_2} B^{(2)} \right)~ , \;\; \ell, m \in \mathbb{Z}~.\label{BFops}
\eeq
Since there are no dynamical degrees of freedom in the theory, there is no dynamical screening of these operators and they are absolutely stable. 

In other words, the operators in eq. \eqref{BFops} are protected by the $\IZ_n$ global symmetries above: the lines are charged under the 1-form global symmetry $\mathbb{Z}_n^{(1)}$ and the surface operators are charged under the 2-form global symmetry $\mathbb{Z}_n^{(2)}$. 
This is seen by checking that these Wilson operators transform under the global symmetries as
\bea
&& \mathbb{Z}_n^{(1)} : \;\; W_1 (\ell, \Sigma_1)~ \longmapsto  e^{\frac{ 2\pi i\ell}{n} \oint_{\Sigma_1} \frac{\epsilon^{(1)}}{2\pi} } W_1 (\ell, \Sigma_1) \\
&& \mathbb{Z}_n^{(2)} : \;\; W_2 (m, \Sigma_2) \longmapsto e^{\frac{ 2\pi i\,m}{n} \oint_{\Sigma_2} \frac{\epsilon^{(2)}}{2\pi} } W_2 (m, \Sigma_2).
\eea
In order to see that the spectrum of these operators are consistent with $\mathbb{Z}_n$, we first recall that $A^{(1)} = \frac{1}{n} d \varphi$. This shows 
\beq
\Big( W_1 (1, \Sigma_1 ) \Big)^n = {\rm exp} \left( i \oint_{\Sigma_1} n A^{(1)} \right) = {\rm exp} \left( i \oint_{\Sigma_1} d \varphi \right) = 1~.
\eeq
Therefore, the Wilson line operators are classified by a charge $\ell = 0, \cdots, (n-1)$,  as we expect for $\mathbb{Z}_n^{(1)}$ symmetry. \\

For the surface operator, we further dualize $A^{(1)}$ to the dual photon field $\tilde{A}^{(1)}$. To this end, we view the field strength $F^{(2)} = d A^{(1)}$ as an independent field and add a Lagrange multiplier term to impose the Bianchi identity
\beq
\mathcal{L} = \frac{in}{2\pi} B^{(2)} \wedge F^{(2)} + \frac{i}{2\pi} d \tilde{A}^{(1)} \wedge F^{(2)} = \frac{i}{2\pi} F^{(2)} \wedge \left( d \tilde{A}^{(1)} + n B^{(2)} \right)~.
\label{eq:app_BF_action_dual}
\eeq
In this presentation, one views the dual photon field as a matter field that Higgses the 2-form gauge field $B^{(2)}$ with charge $n$: $U(1)$ 1-form gauge symmetry is broken down to $\mathbb{Z}_n^{(1)}$. This is linked to the fact that the gauge symmetry of eq.~(\ref{eq:app_BF_action_dual}) is 
\bea
\tilde{A}^{(1)} & \to & \tilde{A}^{(1)} + d \tilde{\lambda}^{(0)} - n \lambda^{(1)} \\
B^{(2)} & \to & B^{(2)} + d \lambda^{(1)}
\eea
where we see that $\tilde{A}^{(1)}$ also comes with its own 0-form gauge symmetry with parameter $\tilde{\lambda}^{(0)}$. This is an emergent gauge symmetry as a result of the duality transformation (similarly to the 1-form gauge symmetry of $B^{(2)}$). The equation of motion for $F^{(2)}$ sets $d \tilde{A}^{(1)} + n B^{(2)} = 0$. This is all we need to prove the $\mathbb{Z}_n$ spectrum of surface operators. For the sake of completeness, we repeat the exercise:
\beq
\left( W_2 (1, \Sigma_2) \right)^n = {\rm exp} \left( i \oint_{\Sigma_2} n B^{(2)} \right) = {\rm exp} \left( i \oint_{\Sigma_2} d \tilde{A}^{(1)} \right) = 1~.
\eeq

In order to understand the correlation function of Wilson operators, we first show that an insertion of surface operator $W_2 (m, \Sigma_2)$ in the path integral (equivalent to introducing a source term) modifies the equation of motion of $B^{(2)}$ and effectively turns on non-trivial $\mathbb{Z}_n$ valued $F^{(2)}$ flux localized on the worldvolume of the surface operator:
\beq
\int \left[ d A^{(1)} \right] \left[ d B^{(2)} \right] \; \underbrace{e^{i m \oint_{\Sigma_2} B^{(2)}}}_{ W_2 (m, \Sigma_2)} e^{\frac{i n}{2\pi} \int_{M_4} B^{(2)} \wedge d A^{(1)} } \quad\Longrightarrow \quad F^{(2)} = -m  \frac{2\pi}{n} \delta^{(2)} (\Sigma_2)~.
\eeq 
Here, $\delta^{(2)} (\Sigma_2)$ is a 2-form delta function which is non-zero only on $\Sigma_2$ and is normalized as $\int_{\Gamma_2} \delta^{(2)} (\Sigma_2) = 1$ where $\Gamma_2$ is a 2-manifold that transversely intersects $\Sigma_2$ once. The modified equation of motion means that the holonomy of $W_1 = e^{i \oint A^{(1)}}$ around $\Sigma_2$ evaluates to a phase $e^{\frac{2\pi i}{n}}$.\footnote{Physically, surface operators $W_2 (m, \Sigma_2)$ are cosmic strings of the UV Abelian Higgs model. This non-trivial holonomy of $W_1$ around the BF string worldvolume is a statement that BF strings are supported by a $\mathbb{Z}_n$ valued magnetic flux.} Similarly, an insertion of a line operator modifies the equation of motion of $A^{(1)}$ to $d B^{(2)} = \frac{2\pi}{n} \delta^{(1)} (\Sigma_1)$, inducing holonomy for $W_2 = e^{i \oint B^{(2)}}$ which $\mathbb{Z}_n$ valued. This means that 
\beq
\langle W_1 (\ell, \Sigma_1) \,W_2 (m, \Sigma_2) \rangle \sim {\rm exp} \left( \frac{2\pi i}{n} \ell m \text{ Link}(\Sigma_1, \Sigma_2) \right)
\eeq 
where $\text{Link}(\Sigma_1, \Sigma_2)$ is the linking number of $\Sigma_1$ and $\Sigma_2$.

Note that the $\mathbb{Z}_n$ theory does not have any non-trivial 't Hooft operators (dual to Wilson operators for $U(1)$ gauge symmetries) nor any local operators (dual of string operators for $U(1)$ symmetries).  The reason is that the local operators constructed from the dual of $B^{(2)}$ are written in terms of $\varphi$:  $I(m, x) = e^{i m \varphi (x)},~m \in \mathbb{Z}$ which is not gauge invariant. However, it can be made gauge invariant by attaching a $\IZ_n$ Wilson line operator to it (see eq.~(\ref{eq:AHM_nonlinear}))
\beq
\tilde{I} (m,x) = e^{i m \varphi (x)} e^{-i m n \int_x A^{(1)}}
\eeq
at the expense that the operator has now become non-local. 
The equation of motion, on the other hand, sets $d \varphi - n A^{(1)} = 0$, showing that the above operator, while gauge invariant, is trivial. 

Similarly, one may consider an 't Hooft line operator $T(\ell, \Sigma_1) = e^{i \ell \oint \tilde{A}^{(1)}}$. The dual formulation eq.~(\ref{eq:app_BF_action_dual}) shows that such an operator is not gauge invariant under $B^{(2)}$ gauge transformations. It can similarly be made gauge invariant by attaching it to a surface operator
\beq
\tilde{T} (\ell, \Sigma_1) = e^{i \ell \oint \tilde{A}^{(1)}} e^{i \ell n \int_{\Sigma_2} B^{(2)}}~,\quad  \partial \Sigma_2 = \Sigma_1
\eeq
Again, the equation of motion makes this a trivial operator.

Before we conclude, we would like to point out that there exist interesting variations of the BF theory by adding a discrete $\theta$-parameter term.
\beq
S = \int_{M_4} \frac{in}{2\pi} B^{(2)} \wedge d A^{(1)} + \frac{ipn}{4\pi} B^{(2)} \wedge B^{(2)}~. 
\eeq
Here, $p \in \mathbb{Z}$. The discrete $\theta$-parameter $\frac{ipn}{4\pi}\int B^{(2)} \wedge B^{(2)}$ is also known as ``discrete torsion'' and theories with different choices of $p$ describe different 
``Symmetry Protected Topological  phases'' (SPT phases) associated with $\mathbb{Z}_n^{(1)}$ symmetry. This theory appears in many situations including in the discussion of $SU(n) / \mathbb{Z}_n$ gauge theory (see for e.g.~\cite{Kapustin:2014gua, Gaiotto:2014kfa, Anber:2021iip}) and non-invertible symmetry as a way of gauging $\mathbb{Z}_n$ subgroup of bulk 1-form magnetic symmetry (see for e.g.~\cite{Choi:2022jqy}). This addition introduces several interesting and subtle effects. It makes the 1-form gauge field $A^{(1)}$ charged under the 1-form gauge symmetry of $B^{(2)}$, it modifies the global symmetry to $\mathbb{Z}_n^{(1)} \times \mathbb{Z}_J^{(2)},$ where $J = \frac{1}{2} {\rm gcd} (p,n)$for $p,n$ even and $J = {\rm gcd} (p,n)$ otherwise. This clearly changes the spectrum of conserved operators and may have interesting signals different from those discussed in our paper. For a very nice and detailed discussion, we recommend highly Section 6 of \cite{Kapustin:2014gua}.

\section{Global Symmetries of TQFT-Coupling I}
\label{app:symmetry_TQFT-coupling I}
 
In this appendix, we discuss in detail symmetries of the theory studied in Section \ref{sec:TQFT-coupling I}: axion-Maxwell theory coupled to a $\IZ_n$ gauge theory.\footnote{ The generalized symmetry and associated higher-group structure of the axion-Maxwell were studied in \cite{Hidaka:2020iaz, Hidaka:2020izy, Brennan:2020ehu}. }

\begin{table}
\begin{tabular}{|c|c|c|c||c|c|c|c|}
\hline
\multicolumn{4}{|c||}{Electric Symmetries} & \multicolumn{4}{c|}{Magnetic Symmetries} \\
\hline\hline 
0-form axion shift & $\mathbb{Z}_{K_A}^{(0)}$ & $* J_{1a}^e$ & $\mathcal{A}_e^{(1)}$ & 2-form axion string & $U(1)^{(2)}$ & $* J_{3a}^m$ & $\mathcal{A}_m^{(3)}$ \\
\hline
1-form $A$-electric & $\mathbb{Z}_{K_A}^{(1)}$ & $ *J_{2A}^e$ & $\mathcal{B}_e^{(2)}$ & 1-form $A$-magnetic & $U(1)^{(1)}$ & $* J_{2A}^m$ & $\mathcal{B}_m^{(2)}$ \\
\hline
1-form $B$-electric & $\mathbb{Z}_{n}^{(1)}$ & $ *J_{2B}^e$ & $\mathcal{C}_e^{(2)}$ &&&&\\
\hline
2-form BF string & $\mathbb{Z}_{n}^{(2)}$ & $* J_{3H}^e$ & $\mathcal{D}_e^{(3)}$ & &&&\\
\hline \hline
\multicolumn{8}{|c|}{Field Strength of Magnetic Symmetries}  \\
\hline \hline
2-form axion string & \multicolumn{3}{c||}{$\mathcal{G}^{(4)} = d \mathcal{A}_m^{(3)} + \cdots$} & 
1-form $A$-magnetic & \multicolumn{3}{c|}{$\mathcal{H}^{(3)} = d \mathcal{B}_m^{(2)} + \cdots$}\\
\hline
\end{tabular}
\caption{List of generalized symmetries, their currents, and background gauge fields in decoupled axion-Maxwell and $\IZ_n$ BF theory. 
}
\label{Tab:list of symm current BGFs}
\end{table}
Since there are many different symmetries and we need to introduce a background gauge field for each of them, we list these symmetries and associated current and background gauge fields in Table~\ref{Tab:list of symm current BGFs}, setting up our notations.\footnote{
Strictly speaking, discrete symmetries do not have associated currents. In Table \ref{Tab:list of symm current BGFs}, by the conserved current generating a discrete symmetry, we mean current of an associated $U(1)$ symmetry that is broken down to the appropriate discrete subgroup. }

The action of the theory is 
\eq{
S = & \frac{1}{2} \int da \wedge * da + \frac{1}{2 g} \int F_A^{(2)} \wedge * F_A^{(2)} + \frac{in}{2\pi} \int B^{(2)} \wedge F_B^{(2)}  \\
- & \frac{i K_A}{8\pi^2 f_a} \int a F_A^{(2)} \wedge F_A^{(2)} - \frac{i K_{AB}}{4\pi^2 f_a} \int a F_A^{(2)} \wedge F_B^{(2)} - \frac{i K_B}{8\pi^2 f_a} \int a F_B^{(2)} \wedge F_B^{(2)}~. 
}
This theory has six generalized global symmetries: four `electric' symmetries and two `magnetic' symmetries. Electric symmetries are obtained from EoMs of the fields appearing in the action and magnetic symmetries come from Bianchi identities.

The procedure for analyzing the symmetries of this theory is to couple the classical symmetries to background gauge fields and study the transformation of the action under background gauge transformations. Then, we can probe the anomalies and higher symmetry structures by coupling all such symmetries to background gauge fields simultaneously and studying their gauge transformations. 

A complimentary viewpoint to study the symmetry groups of the theory is to look at the conservation equation. Take for example the conservation equation associated to a $U(1)$ symmetry. When the symmetry is anomalous, there are extra terms that appear in the conservation equation 
\eq{
d\ast J=j~.
}
When $j$ is a quantized charge density $\int j\in K\IZ$, the $U(1)$ symmetry is broken $U(1)\mapsto \IZ_K$. In a sense, this is a quick and easy way to detect symmetry breaking interactions.

The theory can be coupled to background gauge fields as 
\eq{
S &=  \frac{1}{2} \int (da-f_a \CA_e^{(1)}) \wedge * (da -f_a \CA_e^{(1)})+ \frac{1}{2 g^2} \int (F_A^{(2)}-\CB_e^{(2)}) \wedge * (F_A^{(2)}-\CB_e^{(2)}) \\
+& \frac{in}{2\pi} \int B^{(2)}\wedge (F_B^{(2)}-\CC_e^{(2)}) -\frac{in}{2\pi}\int \CD_e^{(3)}\wedge B^{(1)}
-\frac{i}{2\pi f_a}\int a\,d\CA_m^{(3)}+\frac{i}{2\pi}\int A^{(1)}\wedge d\CB_m^{(2)}\\
- & \frac{i K_A}{8\pi^2 f_a} \int a (F_A^{(2)}-\CB_e^{(2)}) \wedge (F_A^{(2)}-\CB_e^{(2)}) - \frac{i K_{AB}}{4\pi^2 f_a} \int a (F_A^{(2)}-\CB_e^{(2)}) \wedge (F_B^{(2)}
-\CC_e^{(2)}) \\
-& \frac{i K_B}{8\pi^2 f_a} \int a (F_B^{(2)}-\CC_e^{(2)}) \wedge (F_B^{(2)}-\CC_e^{(2)})~. 
}
In general we choose conventions so that all background gauge fields satisfy the analog of Dirac quantization:
\eq{
\oint\frac{d\CA^{(p)}}{2\pi}\in \IZ~. 
}

As we will see below, some of the electric symmetries are mixed up with magnetic symmetries which leads to a higher-group structure. Practically, this happens when the theory only admits field strengths of the magnetic symmetry which are shifted by terms that are non-linear in electric symmetry gauge fields.

\subsection{Symmetries of Uncoupled Theories}
\label{subapp:TQFT-coupling I_GGS}

Now we will discuss the symmetries of the theory. We find that it will be simplest first to analyze the decoupled theory where $K_{AB}=K_B=0$ (i.e. the charge $K_A$ axion-Maxwell theory and $\IZ_{n}$ BF theory are decoupled) starting with the magnetic symmetries before discussing the electric symmetries. We will discuss the symmetry structure of the full theory in the following section.

\subsection*{$U(1)^{(2)}$ 2-form axion string symmetry:}

The axion, being a smooth field obeys a Bianchi identity $d^2 a = 0$. This implies the existence of a two-form $U(1)^{(2)}$ symmetry with a current 
\eq{
* J_3 = \frac{1}{2\pi f_a} d a~.}
This couples to a background 3-form gauge field $\CA_m^{(3)}$
\eq{S =...+ \frac{i}{2\pi f_a} \int  \mathcal{A}_m^{(3)} \wedge d a = i\int  \frac{a}{f_a}  \frac{d \mathcal{A}_m^{(3)}}{2\pi}
}
which is consistent with Dirac quantization condition and the gauge redundancy $a \to a + 2\pi f_a$.

The charged objects of $U(1)^{(2)}$ are 2d world-sheet of axion strings $V (m, \Sigma_2)$. The charge and symmetry defect operators are given by
\eq{
& Q_{aM} (\Sigma_1) = \int_{\Sigma_1} * J_3 ~,\qquad  U_{aM} (\alpha, \Sigma_1) = e^{i \alpha Q_{aM} (\Sigma_1)} .
}
Such symmetry defect operators act on the axion strings with which they have non-trivial linking
\eq{
 \langle U_{aM} (\alpha, \Sigma_1 ) V(m, \Sigma_2) \rangle = e^{i \alpha m \text{ Link} (\Sigma_1, \Sigma_2))} \langle V (m, \Sigma_2) \rangle~.
}
This symmetry is a $U(1)^{(2)}$ and it is broken in the presence of dynamical strings
\beq
d * J_3 = j_2^{\rm string}~.
\eeq
Indeed, in the presence of string, the Bianchi identity $d^2 a = 0$ is violated and the axion carries non-zero winding around the string, see eq.~(\ref{eq:rho_a_delta}). A dual version of this appeared in the BF theory case, i.e.~electric 2-form BF string symmetry discussed in Appendix~\ref{app:review_BF}.

\subsection*{$U(1)^{(1)}$ 1-form $A$-magnetic symmetry:}

This symmetry is a direct consequence of the Bianchi identity $d F_A^{(2)} = 0$ and the current is given by 
\eq{
*J_{2m} = \frac{1}{2\pi} F_A^{(2)}~.}
 If there were dynamical monopoles present in the theory, they would lead to a source term in the conservation equation
\beq
\frac{1}{2\pi} d F_A^{(2)} = j_3^m
\eeq
and this symmetry is broken (possibly with a unbroken discrete subgroup). `t Hooft line operators $T (m, \Sigma_1) = e^{i m \int_{\Sigma_1} \tilde{A}^{(1)}}$ (where $\tilde{A}^{(1)}$ is a dual gauge field) are charged under this magnetic 1-form symmetry. 

This symmetry couples to a background 2-form $U(1)^{(1)}$ gauge field $\CB_m^{(2)}$ as 
\eq{
S=...+\frac{i}{2\pi}\int \CB_m^{(2)}\wedge F_A^{(2)}~. 
}

\subsection*{$\mathbb{Z}_{K_A}^{(0)}$ 0-form axion shift symmetry:}

The equation of motion of the axion is 
\beq
d \left( i f_a * d a \right) = \underbrace{\frac{K_A}{8\pi^2} F_A^{(2)} \wedge F_A^{(2)}}_{= j_4^{\rm inst} (A)}~. 
\label{eq:EoM_a}
\eeq
In the absence of anomalous/instanton terms on the right hand side, the axion has $U(1)^{(0)}$ shift symmetry $a \to a + f_a c, \; c \in \mathbb{S}^1$ and its current is $*J_1 = i f_a * d a$. The above equation shows that the axion coupling breaks the continuous shift symmetry to a discrete subgroup $U(1)^{(0)}\to \IZ_{K_A}^{(0)}$. 

This can also be seen explicitly by studying the shift of the action under $a \to a + f_a c$:
\beq
\delta S = \frac{i c K_A}{8\pi^2} \int F_A^{(2)} \wedge F_A^{(2)} ~.
\eeq
Clearly, the action is only invariant under the shifts
\eq{
a\longmapsto a+\frac{2\pi f_a}{K_A}~,
}
where we used the identity\footnote{Note that this is only true on manifolds with spin structure. On non-spin structure $\int\frac{F_A^{(2)}\wedge F_A^{(2)}}{8\pi^2}\in \half \IZ$, but then there are generically obstructions to having a theory with fermions.}
\eq{
\frac{1}{8\pi^2}\int F_A^{(2)}\wedge F_A^{(2)}\in \IZ~. 
}
We can define a topological symmetry defect operator for $\mathbb{Z}_{K_A}^{(0)}$ by\footnote{In the case with $K_{AB},K_B\neq 0$, the charge operator is modified to 
\eq{
Q_{aE} (\Sigma_3) &= \int_{\Sigma_3} *J_1 -K_A \, \omega_3 (A^{(1)}) - 2 K_{AB} \, \omega_3 (A^{(1)},B^{(1)}) - K_B \, \omega_3 (B^{(1)})
}
where $\omega_3 (A^{(1)},B^{(1)}),\omega_3 (B^{(1)})$ are Chern-Simons terms that satisfy 
\eq{
 d\omega_3 (A^{(1)},B^{(1)})=\frac{1}{8\pi^2}F_A^{(2)}\wedge F_B^{(2)}~,\quad 
 d\omega_3 (B^{(1)})&=\frac{1}{8\pi^2}F_B^{(2)}\wedge F_B^{(2)}~. 
}
}
\eq{
 U_{aE} (\ell, \Sigma_3)= e^{\frac{2\pi i\ell}{K_A} Q_{aE} (\Sigma_3)}~,\quad
Q_{aE} (\Sigma_3) &= \int_{\Sigma_3} *J_1 -K_A \int \omega_3 (A^{(1)}) 
}
where $\ell = 0, 1, \cdots, K_A -1$ and $\omega_3(A^{(1)})$ is a Chern-Simons term that satisfies
\eq{
d\omega_3(A^{(1)})&=\frac{1}{8\pi^2}F_A^{(2)}\wedge F_A^{(2)}~. 
}
The gauge invariant operators that are  charged under this symmetry are the vertex operators $I (m, x) = e^{\frac{i m}{f_a} \,a (x)}, \; m \in \mathbb{Z}$. It is invariant under the gauge transformation $a \to a + 2\pi f_a$ and $U_{aE} (\ell, \Sigma_3)$ acts on it when it has non-trivial linking 
\beq
\langle  U_{aE} (\ell, \Sigma_3) \,I (m, x) \rangle = e^{\frac{2\pi i }{K_a} \ell m \text{ Link} (\Sigma_3, x)} \langle I (m, x) \rangle
\eeq
where on the right-hand side we used the topological invariance of the defect operator $U_{aE} (\ell, V_3)$ to shrink it away after it passes through the charged operator $I(m, x)$.

In order to couple a background gauge field to this symmetry, we in principle need to couple to a $\IZ_{K_A}$ gauge field $\CA_e^{(1)}$. However, since $\IZ_{K_A}\subset U(1)$ we can treat $\CA_e^{(1)}$ as a $U(1)$ gauge field that has been restricted so that:
\eq{
{\rm exp}\left(i\oint \frac{\CA_e^{(1)}}{2\pi}\right)={\rm exp}\left(\frac{2\pi i n}{K_A}\right)~.
}
This has the consequence of imposing the condition 
\eq{
K_A\frac{\CA_e^{(1)}}{2\pi}=d\lambda^{(0)}\sim [0]\in H^1(M_4;\IZ)
}
where $[d\lambda^{(0)}]$ is a trivial cohomology class. Additionally, there is a ``discrete'' version of a $\IZ_n$ gauge field strength which is simply given by\footnote{This operation is called the $\IZ_n$ Bockstein map. }
\eq{
\beta\left(\frac{\CA^{(1)}}{2\pi}\right)=\frac{d\CA^{(1)}}{2\pi}~{\rm mod}_n \in H^2(M_4;\IZ_n)
}
by which we mean $\oint \beta\left(\frac{\CA^{(1)}}{2\pi}\right)=0,1,...,n-1$. See \cite{Kapustin:2014gua} for more in-depth discussion of discrete gauge theories.

There is a mixed `t Hooft anomaly between `electric' $\mathbb{Z}_{K_A}^{(0)}$ shift symmetry and `magnetic' $U(1)^{(2)}$ symmetry. To see this, we couple the theory to background gauge fields
\beq
S =...+ \frac{1}{2} \int \left( d a - f_a \mathcal{A}_e^{(1)} \right) \wedge * \left( d a - f_a \mathcal{A}_e^{(1)} \right) + \frac{i}{2\pi f_a} \int \mathcal{A}_m^{(3)} \wedge \left( d a - f_a \mathcal{A}_e^{(1)} \right)
\eeq
where we added a local counterterm consisting of only background gauge fields to make the action manifestly invariant under electric symmetry. Now we can see that there is a non-trivial `t Hooft anomaly by noting that the action is not invariant under $U(1)^{(2)}$ transformations:
\eq{
\delta S=\frac{i}{2\pi}\int\lambda_m^{(2)}\wedge d\CA_e^{(1)}~,\qquad \delta\CA_m^{(3)}=d\lambda_m^{(2)}~. 
}
 Alternatively, if we chose instead to omit the counterterm, the magnetic symmetry would be preserved but the action would break the axion shift symmetry.
 
 As discussed in Appendix \ref{app:review_GGS}, we can identify this anomaly with a 5D TQFT which is given by 
\beq
S_{\rm inflow} = \frac{i}{2\pi} \int_{N_5} \mathcal{A}_m^{(3)}\wedge d \mathcal{A}_e^{(1)}~.
\eeq
%

\subsection*{$\mathbb{Z}_{K_A}^{(1)}$ 1-form $A$-electric symmetry:}
 
 The 1-form electric symmetry associated to the $U(1)_A$ gauge field is a bit trickier. The equation of motion for $A^{(1)}$ with general $K_{AB}$ is
\beq
d \underbrace{\left(\frac{i}{g_A^2} * F_A^{(2)}\right)}_{=*J_{2A}^e } = \underbrace{- \frac{K_A}{4\pi^2 f_a} da \wedge F_A^{(2)}}_{=j_{3A}^e (a, A^{(1)})} - \underbrace{\frac{K_{AB}}{4\pi^2 f_a} da \wedge F_B^{(2)}}_{= j_{3A}^e (a, B^{(1)})}.
\label{eq:EoM_A}
\eeq
In the absence of coupling terms, the pure Maxwell sector has a $U(1)_e^{(1)}$ 1-form electric symmetry with a current $*J_{2A}^e$.
Naively, the interaction terms would appear to break $U(1)^{(1)}\to \IZ_{K_1}^{(1)}$ for $K_1={\rm GCD}(K_A,K_{AB})$.  

We first consider the case with only $K_A$-term. In that case,  the interaction appears to break  $U(1)_e^{(1)}$ to $\IZ_{K_A}^{(1)}$. To check, let us couple the theory to a 2-form $\IZ_{K_A}$-gauge field $\CB_e^{(2)}$: 
\eq{
S=&... +\frac{1}{2g^2}\int (F_A^{(2)}-\CB_e^{(2)})\wedge \ast (F_A^{(2)}-\CB_e^{(2)})\\
&-\frac{iK_A}{8\pi^2f_a}\int a(F_A^{(2)}-\CB_e^{(2)})\wedge (F_A^{(2)}-\CB_e^{(2)})
}
Since $\CB_e^{(2)}$ is $\IZ_{K_A}$-valued, 
\eq{
\oint \frac{\CB_e^{(2)}}{2\pi}\in \frac{1}{K_A}\IZ \quad \Longrightarrow \quad  \oint \frac{\CB_e^{(2)}\wedge \CB_e^{(2)}}{8\pi^2}\in \frac{1}{K_A^2} \IZ 
}
the coupling is not invariant under the gauge transformation $a \to a +2\pi f_a$ due to the term
\eq{
S = ... + \frac{i K_A}{8\pi^2 f_a} \oint a \CB_e^{(2)}\wedge \CB_e^{(2)}
}
Therefore, we see that the axion interaction term appears to break $U(1)_e^{(1)}\mapsto \IZ_{k_A}^{(1)}$ where $K_A=k\,k_A^2, \;\; k \in \mathbb{Z}$. 

However, we can actually extend the electric 1-form global symmetry to include a $\IZ_{K_A}^{(1)}$ as follows. Recall that the 2-form axion string symmetry $U(1)^{(2)}$ also couples to the axion as
\eq{
S=...-\frac{i}{2\pi f_a}\int a\,d\CA_m^{(3)}~. 
}
If we modify the transformation laws for $\CA_m^{(3)}$, we can cancel the fractional contribution
\eq{&
\delta \mathcal{A}_m^{(3)} =d \lambda_m^{(2)} - \frac{K_A}{4\pi} \left( 2 \lambda_e^{(1)} \wedge \mathcal{B}_e^{(2)} + \lambda_e^{(1)} \wedge d \lambda_e^{(1)} \right)~,\qquad\delta \CB_e^{(2)}=d\lambda_e^{(1)}\label{og3grouptrans1}
}
so that the associated gauge invariant field strength is 
\eq{
\CG^{(4)}=d\CA_m^{(3)}+\frac{K_A}{4\pi}\CB_e^{(2)} \wedge \CB_e^{(2)} ~,
}
and the theory has the coupling 
\eq{
S=&... +\frac{1}{2g^2}\int (F_A^{(2)}-\CB_e^{(2)})\wedge \ast (F_A^{(2)}-\CB_e^{(2)})-\frac{i}{2\pi f_a}\int a\,\CG^{(4)}\\
&-\frac{iK_A}{8\pi^2f_a}\int a(F_A^{(2)}-\CB_e^{(2)})\wedge (F_A^{(2)}-\CB_e^{(2)})~. 
}
We then see that there is a $\IZ_{K_A}^{(1)}$ 1-form global symmetry that mixes with the 2-form axion winding symmetry where $\IZ_{k_A}\subset \IZ_{K_A}^{(1)}$ is a genuine, decoupled 1-form global symmetry. 
This modified transformation law \eqref{og3grouptrans1} is the hallmark of a 3-group global symmetry. 

However, now that the 1-form electric symmetry has mixed with $U(1)^{(2)}$, it also acquires a mixed `t Hooft anomaly with $\IZ_{K_A}^{(0)}$ axion shift symmetry. However, in order to probe these anomalies, we really need to determine how to turn on the background gauge field for $\IZ_{K_A}^{(0)}$ and $\IZ_{K_A}^{(1)}$ simultaneously. This requires picking a 5-manifolds $N_5$ that bounds $M_4$. In this case we find that 
\eq{
S=...+\frac{iK_A}{8\pi^2 f_a}\int_{N_5} (da-f_a \CA_e^{(1)})\wedge (F_A^{(2)}-\CB_e^{(2)})\wedge (F_A^{(2)}-\CB_e^{(2)})~.
}
The requirement that our action describe a well-defined, local 4d theory then becomes demanding that the theory is independent of the choice of $N_5$. One can explicitly check that the action is not independent of this choice due to the terms 
\eq{
S=...+\frac{iK_A}{4\pi^2}\int \CA_e^{(1)}\wedge \CB_e^{(2)}\wedge F_A^{(2)}~. 
}
However, we see that we can actually cancel this variation by also modifying the $U(1)^{(1)}$ magnetic 1-form global symmetry:
\eq{&
\delta \CB_m^{(2)}=d\lambda_m^{(1)}-\frac{K_A}{2\pi}\left(\lambda_e^{(0)}\mathcal{B}_e^{(2)} + \lambda_e^{(1)} \wedge \mathcal{A}_e^{(1)} + \lambda_e^{(0)} d \lambda_e^{(1)} \right)\\
& \delta \CB_e^{(2)}=d\lambda_e^{(1)}~,\quad \delta \CA_e^{(1)}=d\lambda_e^{(0)}
\label{3groupothertrans}
}
so that the gauge invariant field $U(1)^{(1)}$ field strength is given by 
\eq{
\CH^{(3)}=d\CB_m^{(2)}+\frac{K_A}{2\pi}\CA_e^{(1)}\wedge \CB_m^{(2)}
}
and the $U(1)^{(1)}$ coupling now appears as 
\eq{
S=...+\frac{i}{2\pi}\int A^{(1)}\wedge \CH^{(3)}~. 
}
This is the hallmark of a 2-group global symmetry which is then interlaced with the 3-group global symmetry indicated by the mixing of $\IZ_{K_A}^{(1)}$ with $U(1)^{(2)}$.

\subsection*{$\mathbb{Z}_{n}^{(1)}$ 1-form $B$-electric symmetry:}

Now let us turn to the symmetries of $\IZ_n$ BF theory. As we discussed in Appendix~\ref{app:review_BF}, the BF theory has 1-form $\IZ_n^{(1)}$ symmetry. Here, we present an alternative way to see this. The equations of motion for $B^{(1)}$ is given by
\beq
\frac{n}{2\pi} dB^{(2)}=0~.
\label{eq:EoM_B1}
\eeq
This shows that there is a conserved 2-form ``current'', hence a 1-form global symmetry. One notes, however, that the current $*J_2 = \frac{n}{2\pi} B^{(2)}$ is only invariant under large gauge transformations $\frac{B^{(2)}}{2\pi} \mapsto \frac{B^{(2)}}{2\pi} + \Lambda^{(2)}, \; \oint \Lambda^{(2)} \in \IZ$ modulo $n$, which means that the symmetry is broken to $\IZ_n^{(1)}$.
We can couple this symmetry to a 2-form background gauge field by 
\eq{
S=...+\frac{in}{2\pi}\int (F_B^{(2)}-\CC_e^{(2)})\wedge B^{(2)}~. 
}

\subsection*{$\mathbb{Z}_{n}^{(2)}$ 2-form BF string symmetry:}
 
Similarly, the equation of motion for $B^{(2)}$ is the same as in BF theory
\beq
\frac{n}{2\pi} d B^{(1)}=0~.
\label{eq:EoM_B2}
\eeq  
Again, this shows that there is a ``conserved'' 3-form current, hence a 2-form global symmetry.
The current is only gauge invariant mod$_n$ which breaks the naive $U(1)^{(2)}$ electric symmetry down to a $\IZ_n^{(2)}\subset U(1)^{(2)}$. We can again couple the theory to a $\IZ_n^{(2)}$ background gauge field  $\CD_e^{(3)}$ as 
\eq{
S=\frac{in}{2\pi}\int B^{(1)}\wedge (dB^{(2)}-\CD_e^{(3)})~. 
}
Note that this makes it clear that there is no local way we can turn on both $\CC_e^{(2)},\CD_e^{(3)}$ without the action being non-invariant under $\IZ_n^{(1)}\times \IZ_n^{(2)}$ gauge transformations. This is indicative of a mixed `t Hooft anomaly which we will now discuss.

 \subsubsection{Anomalies}

 Now we are able to discuss the `t Hooft anomalies of the decoupled theory. Here we will turn on all of the background fields and analyze the variation of the partition function under the background gauge transformations.  Here we see that there are terms in the action that are explicitly not invariant under the electric symmetries:\
\eq{S=&...-\frac{i}{2\pi f_a} \int_{M_4} a \,\CG^{(4)}+ \frac{i}{2\pi} \int_{M_4}A^{(1)} \wedge \CH^{(3)} \\
&+\frac{in}{2\pi}\int B^{(2)}\wedge (F_B^{(2)}-\CC_e^{(2)})-\frac{in}{2\pi}\int B^{(1)}\wedge \CD_e^{(3)}
}
which leads to the anomalous variation
\eq{
\delta S=...-\frac{i}{2\pi}\int \lambda_e^{(0)} \CG^{(4)}+\frac{i}{2\pi}\int \lambda^{(1)}_e\wedge \CH^{(3)}+\frac{in}{2\pi}\int \left(\tilde\lambda_e^{(2)}\wedge \CC_e^{(2)}-\tilde\lambda_e^{(1)}\wedge \CD_e^{(3)}\right)~
}
where here 
\eq{&
\delta a=f_a \lambda_e^{(0)}~~,\quad \delta A^{(1)}=\lambda_e^{(1)}~,\quad \delta B^{(2)}=\tilde\lambda_e^{(2)}~,\quad \delta \CD_e^{(3)}=d\tilde\lambda_e^{(2)}~,\\&\delta B^{(1)}=\tilde\lambda_e^{(1)}~,\quad \delta \CC_e^{(2)}=d\tilde\lambda_e^{(1)}~.
}
These anomalies can be described by the 5d TQFT: 
\eq{
S_{\rm inflow}=\frac{i}{2\pi}\int \left(\CA_e^{(1)}\wedge \CG^{(4)}+\CB_e^{(2)}\wedge \CH^{(3)}+ n \CD_e^{(3)}\wedge \CC_e^{(2)}\right)~. 
}

\subsection{Symmetries with TQFT Coupling}
\label{subapp:TQFT-coupling I_3-group} 
 
 Now let us consider how turning on the coupling between axion-Maxwell theory and the $\IZ_n$ BF gauge theory, i.e.~$K_{AB},K_B\neq 0$, affects the symmetry structure. The symmetries that are effected are the electric symmetries that shift the fields that participate in  the axion coupling: the 0-form axion shift symmetry and the 1-form electric symmetries.

Conversely, this means that the symmetries that are not effected by turning on the coupling are:
 \begin{itemize}
 \item $U(1)^{(2)}$ 2-form axion string symmetry:
 
Adding the new axionic couplings to the theory does not affect the symmetry structure of $U(1)^{(2)}$. This is evident from the fact that the normal axion coupling does not affect the winding 2-form symmetry of a $U(1)$-valued scalar field. 
 
\item $U(1)^{(1)}_A$ 1-form $A$-magnetic symmetry:
 
Similarly, the new axionic couplings do not affect the magnetic 1-form symmetry of the $U(1)_A$ gauge field. 

\item $\mathbb{Z}_{n}^{(2)}$ 2-form BF string symmetry:

Additionally, the new axionic couplings do not affect the 2-form BF string symmetry.

\end{itemize}

\noindent Now let us consider the modified symmetries in more detail. 

\subsection*{$\mathbb{Z}_{K_a}^{(0)}$ 0-form axion shift symmetry:}

Now the equation of motion of the axion is 
\beq
d \left( i f_a * d a \right) = \underbrace{\frac{K_A}{8\pi^2} F_A^{(2)} \wedge F_A^{(2)}}_{= j_4^{\rm inst} (A)}+ \underbrace{\frac{K_{AB}}{4\pi^2} F_A^{(2)} \wedge F_B^{(2)}}_{= j_4^{\rm inst} (A,B)} + \underbrace{\frac{K_B}{8\pi^2} F_B^{(2)} \wedge F_B^{(2)}}_{= j_4^{\rm inst} (B)}~.  
\label{eq:EoM_a}
\eeq
We now see that there are additional source terms that can further break the $\IZ_{K_A}^{(0)}$ axion shift symmetry down to the subgroup $\IZ_{K_A}^{(0)}\mapsto \IZ_{K_a}^{(0)}$ where $K_a={\rm GCD}(K_A,K_{AB},K_B)$. 

\subsection*{1-form $A$-electric and $B$-electric symmetry}

Now let us consider the effect of adding the coupling $K_{AB}$ to the 1-form electric symmetry for $U(1)_A$ and 1-form electric symmetry for $\IZ_n$. As we saw, the axion coupling for generic $K_A$ broke the symmetry $U(1)^{(1)}$ to a discrete $\IZ_{K_A}^{(1)}$ part of a 3-group. 

When analyzing the role of the $\IZ_n^{(1)}$ BF 1-form electric symmetry, we can effectively treat $F_B^{(2)}$ as the field strength for a $U(1)$ gauge symmetry and $\CC_e^{(2)}$ as a $\IZ_n^{(1)}\subset U(1)_B^{(1)}$ 2-form background gauge field. The reason is that $F_B^{(2)}$ does indeed satisfy the Dirac quantization condition and hence there is no appreciable difference for the purpose of analyzing symmetries. 

Now the symmetry structure of the theory follows straightforwardly from the analysis of the 3-group symmetry structure of axion-Maxwell theory. In particular, the $K_A$ and $K_B$ coupling break $U(1)^{(1)}_A\mapsto \IZ_{K_A}^{(1)}$ and $\IZ_n^{(1)}\mapsto \IZ_{k_B}$ where $k_B={\rm GCD}(K _B,n)$. 

The mixed axion coupling term is a bit more tricky. Let us denote $L={\rm LCM}(K_A,k_B)$. The $\IZ_{K_A}^{(1)},\IZ_{k_B}^{(1)}$ embed into  a $\IZ_L^{(1)}$ enveloping group. We then see that the $K_{AB}$ coupling now breaks $\IZ_L^{(1)}\mapsto \IZ_M^{(1)}$  where $M={\rm GCD}(L,K_{AB})$. Since $L$ is a least common multiple, breaking $\IZ_L^{(1)} \to \IZ_M^{(1)}$ uniquely determines the unbroken $\IZ_{\kappa_A}^{(1)}\subset \IZ_{K_A}^{(1)}$ and $\IZ_{\kappa_B}^{(1)}\subset \IZ_{k_B}^{(1)}$ as 
\eq{\label{kappadef}
&\kappa_A:={\rm GCD}(K_A,M)={\rm GCD}\Big(K_A,\,{\rm GCD}\big({\rm LCM}({\rm GCD}(K_B,n),K_A)\,,K_{AB}\big)\,\Big)
~,\\
&\kappa_B:={\rm GCD}(k_B,M)={\rm GCD}\Big({\rm GCD}(K_B,n)\,,\,{\rm GCD}\big({\rm LCM}({\rm GCD}(K_B,n),K_A)\,,K_{AB}\big)\,\Big)
~. 
}
In our UV complete model in Section \ref{subsec:UV model}, $K_A=1,~K_{AB}=q,$ and $K_B=q^2$ so that $\kappa_A=1$ and $\kappa_B=q$ and there is only a remanining $\IZ_q^{(1)}$ 1-form global symmetry remaining. 

However, when we define the 5d gauge invariant axionic coupling, we also need to cancel the 5-dimensional dependence of the terms 
\eq{
S=&...+\frac{iK_{AB}}{4\pi^2}\int_{N_5}\CA_e^{(1)}\wedge \left(F_B^{(2)}\wedge \CB_e^{(2)}+F_A^{(2)}\wedge \CC_e^{(2)}\right)\\&
+\frac{iK_B}{4\pi^2}\int_{N_5}\CA_e^{(1)}\wedge F_B^{(2)}\wedge \CC_e^{(2)}~.
}
These additionally need to be canceled by modifying the transformations of $\CB_m^{(2)}$ and $\CD_e^{(3)}$ and are accomplished by a straightforward generalization of eq. \eqref{3groupothertrans}. 

Now we see that the coupled theory has a 3-group symmetry involving $U(1)^{(2)},U(1)^{(1)}$, $\IZ_{\kappa_A}^{(1)}$, $\IZ_{\kappa_B}^{(1)}, \IZ_{K_a}^{(0)}$ where the transformation rules are 
\eq{
\delta \mathcal{A}_m^{(3)} =&d \lambda_m^{(2)} - \frac{K_A}{4\pi} \left( 2 \lambda_e^{(1)} \wedge \mathcal{B}_e^{(2)} + \lambda_e^{(1)} \wedge d \lambda_e^{(1)} \right)\\
&-\frac{K_{AB}}{2\pi}\left( \lambda_e^{(1)} \wedge \mathcal{C}_e^{(2)} +\tilde\lambda_e^{(1)}\wedge \CB_e^{(2)}+ \tilde\lambda_e^{(1)} \wedge d \lambda_e^{(1)} \right)\\
&- \frac{K_B}{4\pi} \left( 2 \tilde\lambda_e^{(1)} \wedge \mathcal{C}_e^{(2)} + \tilde\lambda_e^{(1)} \wedge d \tilde\lambda_e^{(1)} \right)~,\\
\delta \CB_m^{(2)} =&d\lambda_m^{(1)}-\frac{K_A}{2\pi}\left(\lambda_e^{(0)}\mathcal{B}_e^{(2)} + \lambda_e^{(1)} \wedge \mathcal{A}_e^{(1)} + \lambda_e^{(0)} d \lambda_e^{(1)} \right)\\
&-\frac{K_{AB}}{2\pi}\left(\lambda_e^{(0)}\mathcal{C}_e^{(2)} + \tilde\lambda_e^{(1)} \wedge \mathcal{A}_e^{(1)} + \lambda_e^{(0)} d \tilde\lambda_e^{(1)} \right)~,\\
\delta \CD_e^{(3)}=&d\tilde\lambda_e^{(2)}-\frac{K_B}{2\pi}\left(d\lambda_e^{(0)}\wedge \mathcal{C}_e^{(2)} + d\tilde\lambda_e^{(1)} \wedge \mathcal{A}_e^{(1)} + d\lambda_e^{(0)} \wedge d \tilde\lambda_e^{(1)} \right)\\
&-\frac{K_{AB}}{2\pi}\left(d\lambda_e^{(0)}\wedge \mathcal{B}_e^{(2)} + d\lambda_e^{(1)} \wedge \mathcal{A}_e^{(1)} + d\lambda_e^{(0)}\wedge  d \lambda_e^{(1)} \right)~,\\
\delta \CA_e^{(1)}=&d\lambda_e^{(0)}~,\qquad\delta \CB_e^{(2)}=d\lambda_e^{(1)}~,\qquad 
\delta \CC_e^{(2)}=d\tilde\lambda_2^{(1)}~,
}
so that the gauge invariant field strengths are given by 
\eq{\label{Fstrengths}
\CG^{(4)}=&d\CA_m^{(3)}+\frac{K_A}{4\pi}\CB_e^{(2)}\wedge \CB_e^{(2)}+\frac{K_{AB}}{2\pi}\CB_e^{(2)}\wedge \CC_e^{(2)}+\frac{K_B}{4\pi}\CC_e^{(2)}\wedge \CC_e^{(2)}\\
\CH^{(3)}=&d\CB_m^{(2)}+\frac{K_A}{2\pi}\CA_e^{(1)}\wedge \CB_e^{(2)}+\frac{K_{AB}}{2\pi}\CA_e^{(1)}\wedge \CC_e^{(2)}
}
while $\CD_e^{(3)}$ is now replaced by its $\IZ_{K_a}^{(0)}\times \IZ_{\kappa_B}^{(1)}$ gauge-invariant form
\eq{\label{CDshift}
\CD_e^{(3)}\to \widetilde\CD_e^{(3)}=\CD_e^{(3)}+\frac{K_{AB}}{3\pi}\CA_e^{(1)}\wedge \CB_e^{(2)}+\frac{K_B}{2\pi}\CA_e^{(1)}\wedge \CC_e^{(2)}~.
}

Here we can clearly see that the effect of coupling the axion-Maxwell theory to the BF TQFT is that the 3-group structure has been dramatically modified. In summary, the effect of the adding the coupling is given by:
\begin{itemize}
\item The axion shift symmetry is reduced $\IZ_{K_A}^{(0)}\mapsto \IZ_{K_a}^{(0)}$ for $K_a={\rm GCD}(K_A,K_B,K_{AB})$,
\item The 1-form $U(1)_A$ eletric symmetry is reduced $\IZ_{K_A}^{(1)}\mapsto \IZ_{\kappa_A}^{(1)}$ where $\kappa_A$ is defined in eq. \eqref{kappadef},
\item The 1-form BF electric symmetry is reduced $\IZ_{n}^{(1)}\mapsto \IZ_{\kappa_B}^{(1)}$ where $\kappa_B$ is defined in eq. \eqref{kappadef},
\item The 1-form BF electric symmetry now participates in a 3-group that mixes with $U(1)^{(2)},U(1)^{(1)}$ magnetic symmetries as well as the electric symmetries $\IZ_{K_a}^{(0)},\IZ_{\kappa_A}^{(1)},\IZ_n^{(2)}$. 
\end{itemize}

\subsubsection{Anomalies}
The full action with all background field strengths turned on is given by 
\eq{
&S =  \frac{1}{2} \int (da-f_a \CA_e^{(1)}) \wedge * (da -f_a \CA_e^{(1)})+ \frac{1}{2 g} \int (F_A^{(2)}-\CB_e^{(2)}) \wedge * (F_A^{(2)}-\CB_e^{(2)}) \\&
+ \frac{in}{2\pi} \int B^{(2)} \wedge (F_B^{(2)}-\CC_e^{(2)})-\frac{in}{2\pi}\int B^{(1)}\wedge \widetilde\CD^{(3)}\\& -\frac{i}{2\pi f_a}\int a\,\CG^{(4)}+\frac{i}{2\pi}\int A^{(1)}\wedge\CH^{(3)}
-  \frac{i K_A}{8\pi^2 f_a} \int a (F_A^{(2)}-\CB_e^{(2)}) \wedge (F_A^{(2)}-\CB_e^{(2)})
\\&
 - \frac{i K_{AB}}{4\pi^2 f_a} \int a (F_A^{(2)}-\CB_e^{(2)}) \wedge (F_B^{(2)}
-\CC_e^{(2)}) 
- \frac{i K_B}{8\pi^2 f_a} \int a (F_B^{(2)}-\CC_e^{(2)}) \wedge (F_B^{(2)}-\CC_e^{(2)})~. 
}
Similarly, we can straightforwardly write down the anomalies of the theory by looking at the non-invariance of the above action under the electric global symmetries:
\eq{
\delta S=...-\frac{i}{2\pi}\int \lambda_e^{(0)} \CG^{(4)}+\frac{i}{2\pi}\int \lambda^{(1)}_e\wedge \CH^{(3)}+\frac{in}{2\pi}\int \left(\tilde\lambda_e^{(2)}\wedge \CC_e^{(2)}-\tilde\lambda_e^{(1)}\wedge \widetilde\CD_e^{(3)}\right)~
}
where here 
\eq{&
\delta a=f_a \lambda_e^{(0)}~~,\quad \delta A^{(1)}=\lambda_e^{(1)}~,\quad \delta B^{(2)}=\tilde\lambda_e^{(2)}~,\quad \delta \CD_e^{(3)}=d\tilde\lambda_e^{(2)}~,\\&\delta B^{(1)}=\tilde\lambda_e^{(1)}~,\quad \delta \CC_e^{(2)}=d\tilde\lambda_e^{(1)}~.
}
These anomalies can be described by the 5d TQFT:
\eq{
S_{\rm inflow}=\frac{i}{2\pi}\int \left(\CA_e^{(1)}\wedge \CG^{(4)}+\CB_e^{(2)}\wedge \CH^{(3)}+\CD_e^{(3)}\wedge \CC_e^{(2)}\right)~. 
}
The full set of background gauge fields and their associated symmetries are summarized in  Table \ref{Tab:list of symm current BGFs2}. \footnote{In Table \ref{Tab:list of symm current BGFs2}, by the conserved current generating a discrete symmetry, we mean current of an associated $U(1)$ symmetry that is broken down to the appropriate discrete subgroup. }

\begin{table}
\begin{tabular}{|c|c|c|c||c|c|c|c|}
\hline
\multicolumn{4}{|c||}{Electric Symmetries} & \multicolumn{4}{c|}{Magnetic Symmetries} \\
\hline\hline 
0-form axion shift & $\mathbb{Z}_{K_a}^{(0)}$ & $* J_{1a}^e$ & $\mathcal{A}_e^{(1)}$ & 2-form axion string & $U(1)^{(2)}$ & $* J_{3a}^m$ & $\mathcal{A}_m^{(3)}$ \\
\hline
1-form $A$-electric & $\mathbb{Z}_{\kappa_A}^{(1)}$ & $ *J_{2A}^e$ & $\mathcal{B}_e^{(2)}$ & 1-form $A$-magnetic & $U(1)^{(1)}$ & $* J_{2A}^m$ & $\mathcal{B}_m^{(2)}$ \\
\hline
1-form $B$-electric & $\mathbb{Z}_{\kappa_B}^{(1)}$ & $ *J_{2B}^e$ & $\mathcal{C}_e^{(2)}$ &&&&\\
\hline
2-form BF string & $\mathbb{Z}_{n}^{(2)}$ & $* J_{3H}^e$ & $\widetilde\CD_e^{(3)}$ & &&&\\
\hline \hline
\multicolumn{8}{|c|}{Field Strength of Magnetic Symmetries}  \\
\hline \hline
2-form axion string & \multicolumn{3}{c||}{$\mathcal{G}^{(4)} = d \mathcal{A}_m^{(3)} + \cdots$} & 
1-form $A$-magnetic & \multicolumn{3}{c||}{$\mathcal{H}^{(3)} = d \mathcal{B}_m^{(2)} + \cdots$}\\ 
\hline
\end{tabular}
\caption{List of generalized symmetries, their currents, and background gauge fields in the full coupled axion-Maxwell and $\IZ_n$ BF theory. Additionally, $\widetilde\CD_e^{(3)}$ is defined in eq.~\eqref{CDshift} and $\CG^{(4)},\CH^{(3)}$ are defined in eq.~\eqref{Fstrengths}, indicating that the 0- and 1-form electric symmetries all participate in 3-groups, mixing into the magnetic symmetries and 2-form BF string symmetry. 
\label{Tab:list of symm current BGFs2}}
\end{table}

\subsubsection{Constraints from Symmetry}

As discussed in \cite{Brennan:2020ehu}, one of the physical consequences of having an EFT with 3-group global symmetry is that any UV completion that gives rise to it must satisfy an inequality of scales at which the different components of the 3-group emerge. In particular, since the 0- and 1-form component symmetries turn on the $2$-form $U(1)^{(2)}$ background gauge fields, we must have the inequality 
\eq{
E_{\rm 2\text{-}form}\gtrsim E_{\rm 1\text{-}form}~.
}
In terms of physical quantities, this is given by 
\eq{
T_{\rm string}\gtrsim m^2_{\psi}
}
where $T_{\rm string}$ is the tension of the axion string and $m_\psi$ is the mass of the lightest charged particle, which must be charged under both $U(1)_A,\IZ_n$ (or $U(1)_A,U(1)_B$ where $U(1)_B$ breaks to $\IZ_n$ at a scale $E_{\IZ_n}\gtrsim m_\psi$) that breaks the 1-form electric symmetries.  
See \cite{Brennan:2020ehu} for further discussion. 

\subsubsection{Other TQFT Couplings via Discrete Gauging}

As discussed in Section \ref{subsec:other TQFT couplings}, we can get many new couplings to TQFTs by gauging discrete subgroups of the 3-group global symmetry. Due to the similarity of the structure of the 3-group, we find that most of the possible discrete gaugings follow straightforwardly. The main difference is that now we can additionally gauge 1.) $\IZ_n^{(2)}$,  2.) $\IZ_{\kappa_B}^{(1)}$, and 3.) $\IZ_{\kappa_A}^{(1)}\times \IZ_{\kappa_B}^{(1)}$. \\

Case 1.) is straightfoward and breaks the $\IZ_{\kappa_B}^{(1)}$ 1-form symmetry due to their mixed anomaly. 

Case 2.) is similar to the case of gauging just $\IZ_{\kappa_A}^{(1)}$ and requires additionally gauging a discrete subgroup of $U(1)^{(2)}$. This additionally breaks $\IZ_{K_a}^{(0)}$ due to an ABJ anomaly and (at least partially) breaks $\IZ_n^{(2)}$. 

Case 3.) combines the effects of gauging $\IZ_{\kappa_A}^{(1)}$ and $\IZ_{\kappa_B}^{(1)}$. It requires gauging a discrete subgroup of $U(1)^{(2)}$, breaks $\IZ_{K_a}^{(0)}$, extends the periodicity of $U(1)^{(1)}$ and (at least partially) breaks $\IZ_n^{(2)}$.

It would be interesting to study the theories produced by these discrete gaugings in more detail.


\bibliographystyle{utphys}
\bibliography{ref}

\end{document}